\pdfoutput=1
\documentclass[conference]{IEEEtran}

\usepackage{url}
\usepackage{comment}
\usepackage{algorithm}
\usepackage{algorithmicx}
\usepackage{algpseudocode}
\usepackage{amsmath}

\usepackage{amssymb}
\usepackage{float}
\usepackage{graphicx} %
\usepackage{booktabs} %
\usepackage[hidelinks]{hyperref}
\usepackage{balance}  %

\newtheorem{definition}{Definition}

\newif\iffullversion
\fullversiontrue %
\newcommand{\fv}[2]{\iffullversion#1\else#2\fi}
\newcommand{\fvonly}[1]{\iffullversion#1\fi}

\usepackage{xcolor}
\definecolor{revyellow}{RGB}{204,153,0}
\definecolor{revred}{RGB}{200,30,30}
\newif\ifreviewhighlight
\reviewhighlightfalse %
\newcommand{\chg}[1]{\ifreviewhighlight{\color{revyellow}#1}\else#1\fi}  %
\newcommand{\chgr}[1]{\ifreviewhighlight{\color{revred}#1}\else#1\fi}   %
\definecolor{revblue}{RGB}{0,90,200}
\newcommand{\chgb}[1]{\ifreviewhighlight{\color{revblue}#1}\else#1\fi}  %
\newif\ifredhighlight
\redhighlightfalse
\newcommand{\chgred}[1]{\ifredhighlight{\color{revred}#1}\else#1\fi}

\AtBeginDocument{%
  }

\begin{document}

\title{jXBW: A Compressed Index Enabling Structure-Aware JSONL Retrieval \\ for Structured RAG}
\author{\IEEEauthorblockN{Yasuo Tabei}
\IEEEauthorblockA{RIKEN Center for Advanced Intelligence Project\\
Tokyo, Japan\\
yasuo.tabei@riken.jp}}

\maketitle

\begin{abstract}
\chg{Providing \textit{structured} information to large language models (LLMs) \chgred{has been shown to improve} multi-step reasoning and factual \chgred{accuracy (e.g., StructRAG, ICLR 2025; RAS, ICLR 2026);} recent retrieval-augmented generation (RAG) systems therefore reconstruct structure from retrieved text on every query. When the corpus is \emph{already} structured --- as in JSON Lines (JSONL), a popular format for LLM prompts, chemical compounds, and geospatial \chgred{objects} --- this per-query rebuilding can be replaced by direct \emph{structural retrieval}. The core \chgred{operation of structural retrieval} is \textit{substructure search}: finding all JSON objects in a collection that contain a given query \chgred{tree}. Existing approaches index each \chgred{object} separately, so both index space and query time grow with the total collection size; XML-based engines add conversion overhead and semantic mismatches. We propose \textbf{jXBW}, a compressed index for fast substructure search over JSONL, combining three innovations: (i) a merged tree representation that consolidates repeated structures across objects, (ii) a succinct tree \chgred{encoding} based on the eXtended Burrows--Wheeler Transform (XBW), and (iii) a three-phase substructure search algorithm on this \chgred{encoding}. Together they achieve \textbf{query-dependent complexity} \chgred{in compressed space}: \chgb{the search avoids a full scan of the collection\chgred{, and its cost grows with the numbers of candidate matches and results}}. Experiments on seven real-world datasets, including PubChem ($10^6$ compounds) and OpenStreetMap ($6.6 \times 10^6$ objects), show that jXBW outperforms the strongest tree-based baseline by $\mathbf{16\times}$ on the smallest dataset and by up to $\mathbf{2{,}800\times}$ on the largest, \chgred{while} the \chgred{XML-based} engine Saxon \chgred{does not complete the $1{,}000$-query workload on PubChem within $24$\,hours}. jXBW brings \chgred{substructure search} over \chgred{million-object} JSONL collections into the sub-millisecond range.}

\end{abstract}


\section{Introduction}

Providing \emph{structured} information to large language models (LLMs), rather than only flat text, has been shown to improve multi-step reasoning and factual \chgred{accuracy} in knowledge-intensive tasks. Two recent results make this concrete\chgred{.} StructRAG (ICLR 2025)~\cite{li2025structrag} \emph{reconstructs} retrieved documents into task-appropriate structures (tables, graphs, trees) at inference time and reports state-of-the-art accuracy on knowledge-intensive benchmarks\chgred{.} Retrieval-And-Structuring (RAS, ICLR 2026)~\cite{jiang2026ras} \emph{dynamically builds} question-specific knowledge graphs by interleaving retrieval with graph construction \chgred{and gains} up to $8.7\%$ over standard retrieval-augmented generation (RAG)~\cite{lewis2020rag}.
Both systems \chgred{pay for} this \chgred{gain by} rebuilding \chgred{the} structure on every query, because their source corpora are unstructured text.

When the source corpus is \emph{itself} already in structured form, this per-query rebuilding can in principle be replaced by direct \emph{structural retrieval}. A particularly common substrate is \emph{JSON Lines} (JSONL): a format storing massive collections of nested, self-contained JSON objects, one per line. JSON underlies structured LLM I/O in real-world applications~\cite{wang2024systematic}, and JSONL extends this to collection-scale corpora: public scientific repositories (PubChem: $10^6$ compounds), geospatial datasets (OpenStreetMap\chgred{:} $> 10^6$ objects per region), and prompt or instruction corpora for LLM workflows~\cite{chen2024genqa}. \chgb{However, no existing system \chgred{retrieves such objects efficiently by their structure} (Section~\ref{sec:related-work}).}
Thus, developing efficient structural retrieval at JSONL scale remains an important open challenge.

Queries over such collections \chgred{are not captured} by keyword or text-similarity \chgred{matching alone} --- they describe a \emph{structural fragment} that the target \chgred{object} must contain (a nested chemical-attribute fragment in a PubChem record, a nested-attribute pattern on an \chgred{OpenStreetMap object}, a schema fragment in a prompt corpus). 
We model each JSON object \chgred{in a JSONL collection} as a labeled tree where each internal node has either unordered children (an \emph{object}: a set of key-value pairs) or ordered children (an \emph{array}: a positional list of values), and leaves carry primitive values. Under this representation, the corresponding retrieval \chgred{operation} is \emph{substructure search}: given a query \chgred{tree built in the same way}, return every \chgred{object} whose tree contains it as a \chgred{substructure}.

\chgb{However, a substructure search query cannot be answered by standard ordered-tree or unordered-tree pattern matching alone: an ordered matcher would wrongly require object keys to appear in a fixed order, while an unordered matcher would wrongly ignore array element order.} The matcher must respect both regimes simultaneously. \chgb{Prior structural approaches\chgred{, moreover, index} each \chgred{object} independently.} \chgred{Per-tree} traversal and succinct representations (SJSON) match a query against every tree in turn; \chgb{Saxon evaluates an XQuery expression per \chgred{object}}, taking \chgred{about $2$\,s} per query on \chgred{$6.6\times 10^6$} \chgred{objects} (Section~\ref{sec:experiments}). Across all prior approaches, both index space and query time scale with $\sum_i |T_i|$ or $N$, where $N$ is the number of \chgred{objects} in the collection and $|T_i|$ is the number of nodes in tree \chgred{$T_i$} (Table~\ref{tab:complexity}).

\begin{table*}[t]
\centering
\caption{\chgr{Index space (in bits) and query time for substructure search over a JSONL collection of $N$ \chgred{objects} with trees $T_i$. $|MT| \leq \sum_i |T_i|$: merged-tree size\chgred{; $MT'$: $MT$ with normalized labels (Section~\ref{sec:jxbw_structure})}\chgred{; $|MT'|$: size of $MT'$}; $\sigma$: label alphabet; $|Q|$: query size; $p$, $d$: number and average depth of root-to-leaf query paths; $r$: matching leaf positions; $c$: \chgred{candidate roots}; $L$: total symbol-table label length; $|A_{\mathit{ids}}|$: stored tree identifiers; \chgb{$I_{\mathit{ids}}$: identifier entries examined during tree-identifier intersection and output collection; $V$: verification cost, zero for array-free queries;} $\tau_{\mathit{XQ}}$: \chgred{per-object} XQuery evaluation cost. \chgb{DuckDB's bound reflects the configuration in Section~\ref{sec:experiments}: no persistent structural index, so every query requires a full unindexed containment scan.}}}
\label{tab:complexity}
\small
\begin{tabular}{lllll}
\toprule
Method & Index strategy & Index type & Index space & Query time \\
\midrule
\multicolumn{5}{l}{\emph{Prior approaches:}} \\
Per-tree traversal & \chgred{Per-object} & Pointer & \chgr{$O(\sum_i |T_i| \log \sum_i |T_i|)$} & $O(\sum_i |T_i| \cdot |Q|)$ \\
SJSON~\cite{lee2020sjson} & \chgred{Per-object} & LOUDS (succinct) & \chgr{$O(\sum_i |T_i| \log \sigma)$} & $O(\sum_i |T_i| \cdot |Q|)$ \\
\chgb{DuckDB}~\cite{raasveldt2019duckdb} & \chgred{Per-object} & No index & $O(\sum_i |T_i|)$ & $O(\sum_i |T_i| \cdot |Q|)$ \\
Saxon~\cite{Kay08,saxon_processor} & \chgr{\chgred{Per-object}} & XML index & \chgr{$O(\sum_i |T_i| \log \sum_i |T_i|)$} & $\Omega(N \cdot \tau_{\mathit{XQ}})$ \\
\midrule
\multicolumn{5}{l}{\emph{This work (merged-tree variants, sharing the merged tree of \S\ref{sec:merged_tree}):}} \\
Ptree & Merged tree & Pointer & \chgr{\begin{tabular}[t]{@{}l@{}}$O(|MT| \log |MT| + L$\\ $\;\;+ |A_{\mathit{ids}}| \log N)$\end{tabular}} & $O(|MT| \cdot |Q|)$ \\
SucTree & Merged tree & LOUDS (succinct) & \chgr{\begin{tabular}[t]{@{}l@{}}$O(|MT| \log \sigma + L$\\ $\;\;+ |A_{\mathit{ids}}| \log N)$\end{tabular}} & $O(|MT| \cdot |Q|)$ \\
\textbf{jXBW} (proposed) & Merged tree & \chgb{XBW (succinct)} & \begin{tabular}[t]{@{}l@{}}$O(|MT'| \log \sigma + L$\\ $\;\;+ |A_{\mathit{ids}}| \log N)$\end{tabular} & \begin{tabular}[t]{@{}l@{}}\chgred{$O((p+r) \cdot d \cdot \log \sigma$}\\ $\;\;+ c \cdot p \cdot d \cdot \log \sigma \chgb{{} + I_{\mathit{ids}} + V})$\end{tabular} \\
\bottomrule
\end{tabular}
\end{table*}

{\em Contribution.} \chgred{We} present \chgred{\emph{jXBW},} a new compressed index for substructure search over large-scale JSONL collections. A key idea is to consolidate the structural redundancy across all \chgred{objects} into a single \emph{merged tree} (Section~\ref{sec:merged_tree}) while preserving the identity of each \chgred{object}. To support fast search on this representation, we propose a novel succinct encoding of the merged tree based on the eXtended Burrows--Wheeler Transform (XBW)~\cite{ferragina2007xbw} that supports tree navigation and subpath search in compressed space. On top of this encoding, we design a new three-phase substructure search algorithm \chgred{that matches the query paths, finds candidate roots, and collects tree identifiers adaptively} (Section~\ref{sec:substructure_search_json}). jXBW has the following properties:
\begin{enumerate}
  \item \textbf{Scalability:} jXBW is applicable to massive JSONL collections; its index scales with the merged-tree size, which shrinks as \chgred{objects} share structure.
  \item \textbf{Query-dependent complexity:} query time is governed by \chgred{the} query structure and the \chgred{numbers} of \chgred{candidate roots} and \chgred{matches} (Table~\ref{tab:complexity}), rather than by a linear scan over the collection.
  \item \textbf{\chgred{Space efficiency}:} \chgb{the main search operations run directly on the compressed XBW representation}, with memory usage competitive with succinct baselines and below pointer-based ones.
  \item \textbf{Generality:} \chgb{jXBW supports both ordered (array) and unordered (object) semantics}, making \chgred{it} applicable to \chgred{a wide range of} JSONL \chgred{collections} --- from chemical compounds to geospatial objects to LLM instruction data.
\end{enumerate}
We experimentally test jXBW on seven real-world datasets with up to $6.6 \times 10^6$ objects, and show that it \chg{outperforms the strongest tree-based baseline (Ptree) by $16\times$ on the smallest dataset and by up to $2{,}800\times$ on the largest}\chgred{. On all seven datasets}, \chgred{it answers queries in the sub-millisecond range; on PubChem ($10^6$ compounds),} the XML-based engine Saxon \chgb{does not complete the $1{,}000$-query workload within $24$\,hours}.
\chgb{Code, data, and experimental scripts are publicly available at \url{https://github.com/tb-yasu/jXBW}.}

\section{Related Work}
\label{sec:related-work}

\chgred{This section reviews substructure search over JSONL, and Table~\ref{tab:complexity} summarizes state-of-the-art approaches.}

\paragraph{XML pattern matching and XPath/XQuery engines}
Tree-pattern queries over semi-structured data have a long history in the XML community. XPath/XQuery~\cite{W3C2017XQuery3_1} engines such as Saxon~\cite{Kay08,saxon_processor} and BaseX implement mature indexing for path expressions and structural joins~\cite{BrunoKoudasSrivastava2002,RautAtique2014}. \chgred{One can convert each JSON object} to XML \chgred{and query it with} these engines (as our Saxon baseline does; Section~\ref{sec:experiments})\chgred{, but the engines still evaluate the query on every object separately}\chgred{. In addition}, \chgred{this} conversion is costly: JSON arrays, \texttt{null} values, and primitive types have no direct XML counterparts and must be encoded with extra markup.

\chgb{\paragraph{JSON stores and analytical engines}

Mainstream \chgred{data} systems support JSON natively\chgred{.} PostgreSQL stores \chgred{objects} as
\texttt{jsonb} \chgred{and indexes them} with \chgred{generalized inverted} indexes \chgred{(GIN)} for containment and JSON-path predicates\chgred{;}
MongoDB indexes nested BSON \chgred{(binary JSON)} field paths and array elements\chgred{;}
DuckDB~\cite{raasveldt2019duckdb} provides vectorized JSON extraction and
containment\chgred{;} and relational approaches such as Argo~\cite{chasseur2013argo}
map \chgred{JSON objects} into relational \chgred{tables}. \chgred{All of them} are designed for \chgred{predicates on the fields}, \chgred{paths}, or \chgred{values of individual objects, not} for \chgred{substructure search: finding all objects in a collection that contain} a query tree at any position.}

\paragraph{Succinct tree representations}

A succinct data structure~\cite{jacobson1989space,navarro2016compact} is a compressed data structure that uses space close to the information-theoretic lower bound while supporting efficient queries.
SJSON~\cite{lee2020sjson} represents JSON documents with a succinct tree encoding based on LOUDS (level-order unary degree sequence), but targets storage compression rather than search acceleration. None of these structures exploits the structural redundancy \emph{across} the \chgred{objects} of a JSONL collection. 

\paragraph{Structured retrieval for retrieval-augmented generation}
\chgred{A recent line of work on retrieval-augmented generation (RAG)~\cite{lewis2020rag} provides LLMs with structured context instead of flat text.} StructRAG (ICLR 2025)~\cite{li2025structrag} reconstructs retrieved documents into task-appropriate hybrid structures (tables, graphs, trees) at inference time and reports state-of-the-art accuracy on knowledge-intensive benchmarks\chgred{, and} Retrieval-And-Structuring (RAS, ICLR 2026)~\cite{jiang2026ras} retrieves text passages and \chgred{builds} question-specific knowledge graphs, gaining up to $8.7\%$ over standard RAG. \chgred{In both, the source corpus is \emph{unstructured} text: the structure is built from the retrieved text at inference time and must be rebuilt for every query.} \chgred{jXBW targets the opposite setting, in which the source corpus is \emph{already} structured (JSONL); substructure search then retrieves the relevant objects directly, and no structure is built at query time.} The two directions are complementary\chgred{:} jXBW could serve as the retrieval substrate \chgred{of} a structured RAG pipeline \chgred{over JSONL}, while text-to-structure systems remain necessary \chgred{for} unstructured \chgred{corpora}.

\chgb{Despite the importance of fast substructure search over large-scale JSONL collections, no previous work achieves this goal. Existing approaches either index individual \chgred{objects}, fields, or paths rather than sharing structure across the collection, or they compress each \chgred{object's} tree individually without accelerating substructure search. We present jXBW, an index for fast substructure search built from a single succinct XBW\chgred{~\cite{ferragina2007xbw}} encoding over a merged tree spanning the whole collection. Details of the proposed method are presented in the next section.}

\section{Method}
\chgred{This section defines the tree representation of a JSONL collection and the substructure search problem that jXBW solves.}

\subsection{JSONL and its Tree Representation}
JSONL is a text-based format in which each line \chgred{holds one} JSON object.  
Let a JSONL \chgred{collection} consist of $N$ objects denoted by $O_i$, where $i = 1, 2, \ldots, N$.
\chg{Each JSON object consists of an unordered set of key-value pairs, where each value is a string, a number, a nested object, an ordered array of values, \texttt{true}, \texttt{false}, or \texttt{null}; \fv{Appendix~\ref{sec:json_bnf} gives the precise definition in Backus--Naur Form (BNF)}{the precise definition in Backus--Naur Form (BNF) is given in \chgb{the extended version~\cite{tabei2026jxbwfull}}}.}

An example involving two JSON objects that manage information about two individuals, Alice and Bob, is given as follows:

\begin{minipage}{0.95\linewidth}
\begin{ttfamily}
O$_1$ = \{"person":\{"name":"Alice","age":30\}, "hobbies":["reading","cycling"]\} \\
O$_2$ = \{"person":\{"name":"Bob","age":30\},   "hobbies":["reading"]\}
\end{ttfamily}
\end{minipage} 

\begin{figure}[t]
  \centering
  \includegraphics[width=1.0\linewidth]{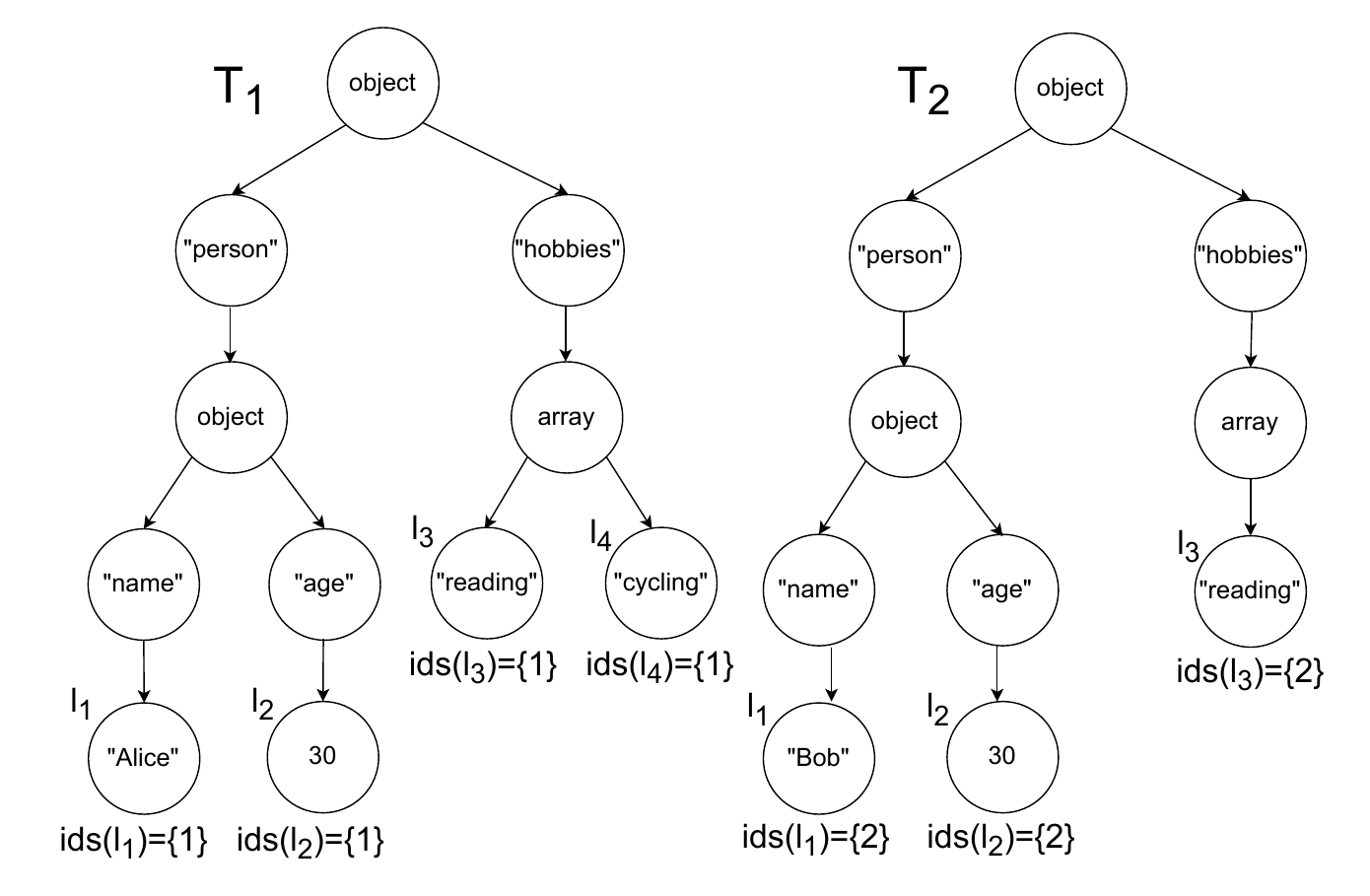}
  \vspace{-0.5cm}
  \caption{Tree structures $T_1$ (left) and $T_2$ (right), derived from the example JSON objects $O_1$ and $O_2$.}
  \label{fig:tree}
\end{figure}
\begin{figure}[t]
  \centering
  \includegraphics[width=0.9\linewidth]{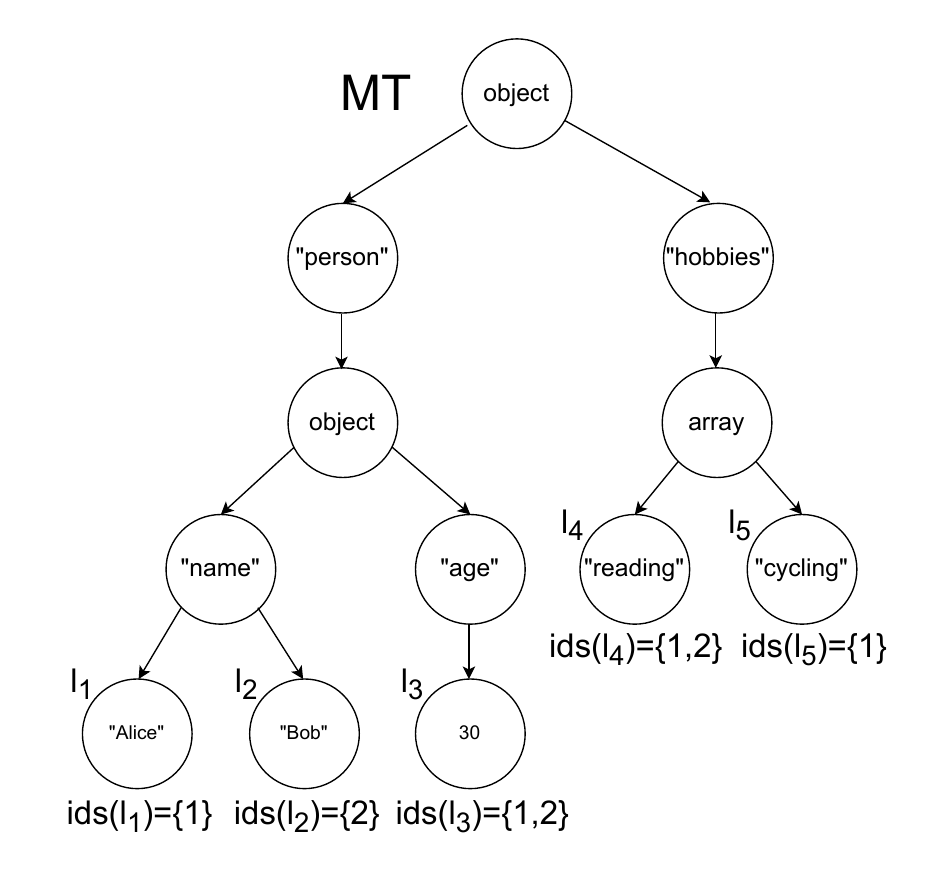}
  \vspace{-0.5cm}
  \caption{Merged tree $MT$ derived from $T_1$ and $T_2$.}
  \label{fig:merged_tree}
\end{figure}

\chg{{\bf Tree Representation.} \chgred{We transform each} object $O_i$ into a labeled tree $T_i$ \chgred{with} $|T_i|$ nodes \chgred{as follows}.
The root of $T_i$ is labeled ``object'' and has one unordered child per key-value pair of $O_i$ (no children if $O_i$ is empty).
Each such child is labeled with the key and has a single child node representing the value.
If the value is a nested object, the value node \chgred{becomes} the root of a \chgred{recursively built} subtree\chgred{.}
\chgred{If} the value is an array, the value node has one child per array element\chgred{,} in \chgred{array} order\chgred{;} each element may again be of any JSON type\chgred{.}
\chgred{Otherwise}, the value node is a leaf carrying the primitive value (a string, a number, \texttt{true}, \texttt{false}, or \texttt{null}).}

\chgr{\chgred{We annotate} each leaf node $\ell$ \chgred{of} $T_i$ with the tree identifier $i$ --- the index of its source object $O_i$ --- so that search results can be traced back to the original JSON objects \chgred{once the trees are merged (Section~\ref{sec:merged_tree})}.}
For any leaf node $\ell$, we define \chgred{the identifier set} $\mathrm{ids}(\ell)$ as the set of all tree identifiers associated with \chgred{$\ell$}.
In \chgred{a single tree $T_i$}, $\mathrm{ids}(\ell) = \{i\}$ for \chgred{every} leaf node $\ell$.

Figure~\ref{fig:tree} \chgred{shows} the \chgred{trees} $T_1$ and $T_2$ \chgred{for} the example objects $O_1$ and $O_2$\chgred{; their} leaf nodes \chgred{carry the tree} identifiers $1$ and $2$, respectively.

\chgred{We formalize} substructure search as follows.
\begin{definition}[Substructure Search Problem]
  \label{def:substructure_search_problem}
  Consider a set of $N$ trees $T_1, T_2,\ldots, T_N$, where each tree $T_i$ is derived from \chgr{the JSON object $O_i$ of a JSONL \chgred{collection}}.
  \chgred{Given} a query tree $Q$, \chgred{the substructure search problem is to} find all \chgred{tree identifiers} $i \in \{1, 2, \ldots, N\}$ such that $T_i$ contains $Q$ as a substructure.
  Here, $Q$ is a substructure of $T_i$ if there exists a mapping from each node in $Q$ to a node in $T_i$ such that (i) corresponding nodes have the same label, (ii) the parent-child relationships are preserved, and (iii) the ordering constraints are satisfied according to the JSON semantics: unordered matching for JSON object children and ordered matching for JSON array children.
  \end{definition}

  \chg{The query tree $Q$ can be derived from any valid JSON structure: \chgred{an object}, \chgred{an array}, or \chgred{a} nested \chgred{combination of both}; \chgred{matching follows} the ordering \chgred{rules} of Definition~\ref{def:substructure_search_problem}.}

A straightforward approach to substructure search in \chgred{Definition}~\ref{def:substructure_search_problem} would involve searching each individual tree $T_i$ separately for the query tree $Q$. 
However, this approach has significant computational drawbacks. First, it requires $O\!\big((\sum_{i=1}^{N} |T_i|)\cdot |Q|\big)$ time complexity, where $|T_i|$ is the number of nodes in $T_i$ and each tree must be traversed independently to find potential matches.
Second, this approach fails to exploit structural similarities between trees, resulting in redundant computations when many trees share common substructures. To overcome 
these limitations, we introduce an efficient merged tree representation that consolidates shared structural patterns while preserving individual tree identities.

\section{Merged Tree Representation}
\label{sec:merged_tree}

\chgr{To address these limitations, we consolidate the $N$ trees into a single \emph{merged tree} $MT$ that stores every distinct root-path \chgred{(the sequence of node labels on the path from the root to a node)} only once.
The merging rule is purely path-based: nodes $u \in T_i$ and $v \in T_j$ are merged into a single node of $MT$ exactly when their \chgred{root-paths} coincide.
Consequently, trees \chgred{whose root-paths share} a prefix share the corresponding nodes of $MT$, and where the \chgred{root-paths} diverge, each divergent child starts a new branch under the last shared node.
Each leaf $\ell$ of $MT$ records the identifier set $\mathrm{ids}(\ell)$, defined as the set of \chgred{tree identifiers} $i$ such that $T_i$ contains a leaf \chgred{with the same} root-path \chgred{as} $\ell$.} \fv{Algorithm~\ref{alg:tree_merge} in Appendix~\ref{sec:merge_trees} \chgred{gives} pseudo-code for tree merging.}{Pseudo-code for tree merging is \chgred{given} in \chgb{the extended version~\cite{tabei2026jxbwfull}}.}

\chgred{Figure~\ref{fig:merged_tree} shows the merged tree of $T_1$ and $T_2$.} 
In this example, both the leaf nodes labeled ``30'' and ``reading'' \chgred{lie on root-paths} shared by $T_1$ and $T_2$, resulting in these leaf nodes $\ell_3$ and $\ell_4$ having $\mathrm{ids}(\ell_3) = \{1, 2\}$ and $\mathrm{ids}(\ell_4) = \{1, 2\}$ in the merged tree. The other leaf nodes $\ell_1$ labeled ``Alice'', $\ell_2$ labeled ``Bob'', and $\ell_5$ labeled ``cycling'' have $\mathrm{ids}(\ell_1) = \{1\}$, $\mathrm{ids}(\ell_2) = \{2\}$, and $\mathrm{ids}(\ell_5) = \{1\}$, respectively.

The merged tree representation provides significant memory efficiency compared to storing individual trees separately. 
The space complexity of the merged tree is $O(|MT|)$, where $|MT| \leq \sum_{i=1}^{N} |T_i|$ with equality only when no structural patterns are shared among the input trees.
In practice, when input trees share common structural patterns—which is typical in real-world JSONL \chgred{collections}—the merged tree achieves substantial space savings with $|MT| \ll \sum_{i=1}^{N} |T_i|$.

Merging \chgred{the} $N$ trees \chgred{one by one takes} $O(M_{\mathrm{tot}}^2)$ time in the worst case, where $M_{\mathrm{tot}} = \sum_{i=1}^{N} |T_i|$ is the total number of input nodes\chgred{. We} therefore merge \chgred{them} with a balanced divide-and-conquer strategy, which reduces the overall cost to $O(M_{\mathrm{tot}} \log N)$ (\fv{the cost analysis is given in Appendix~\ref{sec:merge_trees}}{the cost analysis is given in \chgb{the extended version~\cite{tabei2026jxbwfull}}}).

\subsection{Substructure Search on Merged Tree}
\label{sec:substructure_search_mt}

\begin{figure}[t]
  \centering
  \includegraphics[width=0.55\linewidth]{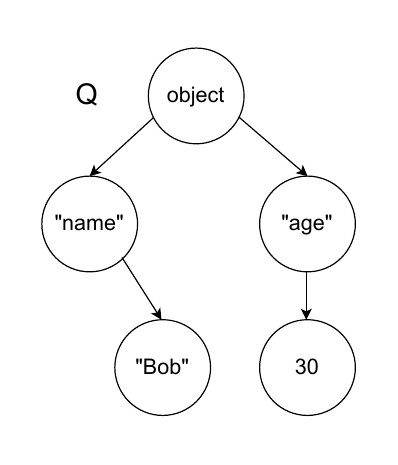}
  \vspace{-0.7cm}
  \caption{Example query tree \chgred{$Q$ for the object $\{\text{"name"}:\text{"Bob"}, \text{"age"}:30\}$}.}
  \label{fig:query}
\end{figure}

\chgred{Because $MT$ merges identical root-paths, one search over $MT$ covers all $N$ trees, in three steps:}
\begin{enumerate}
\item \textbf{Candidate finding} \chgred{selects every node $v$ of $MT$ whose label equals the root label of $Q$.}
\item \textbf{Substructure matching} \chgred{checks recursively that the subtree rooted at $v$ contains $Q$, mapping each leaf of $Q$ to a leaf $\ell$ of $MT$.} This requires exhaustively comparing the labels of child nodes between corresponding nodes in $MT$ and $Q$ to find matches, resulting in computational overhead that is proportional to the product of the branching factors of both the merged tree and query tree nodes.
\item \textbf{Solution identification} \chgred{intersects the sets $\mathrm{ids}(\ell)$ of these matched leaves: since $\mathrm{ids}(\ell)$ lists the trees that have a leaf with the same root-path as $\ell$, every tree containing $Q$ at $v$ lies in all of them; the union over all candidates $v$ is the answer.}
\end{enumerate}

\chgred{Detailed merged-tree} algorithms are \chgred{given} in \fv{Appendix~\ref{sec:mt_algorithms}}{\chgb{the extended version~\cite{tabei2026jxbwfull}}}.

For the example query object $\{\text{"name"}:\text{"Bob"}, \text{"age"}:30\}$, we convert this object into the tree structure $Q$ as shown in Figure~\ref{fig:query}.
The algorithm searches for nodes labeled ``object'' by recursively traversing $MT$ in Figure~\ref{fig:merged_tree} during the candidate finding phase.
It finds root and internal nodes labeled ``object'' in $MT$, and recursively checks whether the subtree rooted at each node matches the structure of $Q$ during the substructure matching phase.
One subtree rooted at an internal node labeled ``object'' is found to \chgred{contain} $Q$.
\chgr{The sets of identifiers $\mathrm{ids}(\ell_2) = \{2\}$ and $\mathrm{ids}(\ell_3) = \{1,2\}$ associated with the leaves $\ell_2$ and $\ell_3$ in $MT$ are collected.
In the solution identification phase, we compute the intersection of all collected identifier sets as $\mathrm{ids}(\ell_2) \cap \mathrm{ids}(\ell_3) = \{2\} \cap \{1,2\} = \{2\}$. Thus the \chgred{answer is} $\{2\}$.}

\textbf{Complexity.} The substructure search algorithm on the merged tree has significant efficiency limitations. The time complexity is \chgred{$O(|MT| \cdot |Q|)$} in the worst case. 
This occurs when the algorithm must examine every node in $MT$ as a potential starting point for matching the root of $Q$, and for each candidate node, recursively compare the entire structure of $Q$ against the corresponding subtree in $MT$.
Furthermore, during the substructure matching phase, for each pair of nodes (one from the substructure in $MT$ and one from $Q$), the algorithm must exhaustively search through all children of the $MT$ node to find those with matching labels to the children of the $Q$ node. 
Although this approach is more efficient than individual searches on each of the $N$ original trees (which would require $O\!\big((\sum_{i=1}^{N} |T_i|)\cdot |Q|\big)$ time\chgred{)}, the need to traverse the entire merged tree remains a computational bottleneck, especially for large JSONL \chgred{collections} with complex tree structures.
\chgred{The merged-tree variants Ptree and SucTree (Table~\ref{tab:complexity}) implement this search: shared prefixes are traversed only once, but each query still scans $MT$, at a worst-case cost of \chgred{$O(|MT| \cdot |Q|)$}.} \chgred{jXBW (Section~\ref{sec:substructure_search_json}) avoids this full scan.}

\section{jXBW}
\begin{figure*}[t]
  \centering
  \includegraphics[width=1.0\linewidth]{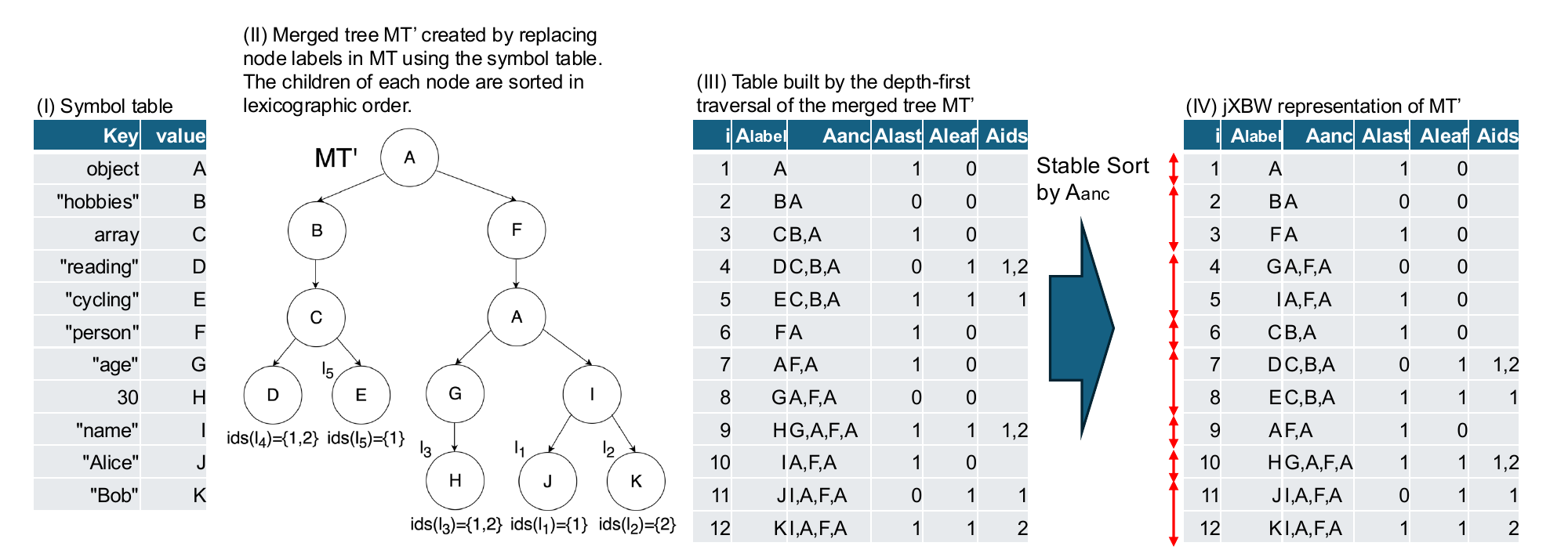}
  \vspace{-0.3cm}
  \caption{The jXBW data structure (panel (IV)) and its construction from the merged tree $MT$: (I) symbol table; (II) normalized merged tree $MT'$; (III) arrays extracted by a depth-first traversal of $MT'$; (IV) the final jXBW arrays, stably sorted by the ancestor label sequence $A_{\mathit{anc}}$. Red brackets in (IV) mark the contiguous sibling blocks (property (i) in Section~\ref{sec:jxbw_structure}). $A_{\mathit{anc}}$ is auxiliary\chgred{: it is shown for clarity but} not stored in the final representation.}
  \label{fig:xbw_construction}
\end{figure*}

The eXtended \chgred{Burrows--Wheeler} Transform (XBW)~\cite{ferragina2007xbw} is a compressed data structure for labeled trees.
XBW \chgred{linearizes a} tree while preserving \chgred{its topology}, \chgred{which enables} (i) efficient navigation\chgred{,} such as moving from a parent to its children and \chgred{back,} and (ii) subpath search\chgred{, which} can locate matching paths \chgred{that start at} any node.
We \chgred{call} the XBW representation of $MT$ \chgred{jXBW; it encodes the} merged tree \chgred{compactly and} supports fast substructure search without explicitly storing the entire tree structure.

\subsection{The jXBW Data Structure}
\label{sec:jxbw_structure}
jXBW \chgred{combines} a symbol table \chgred{with} synchronized arrays \chgred{(}Figure~\ref{fig:xbw_construction}\chgred{)}.
\chgred{The node labels of} $MT$ \chgred{come} from the original \chgred{JSON} objects \chgred{and include field names} (e.g., ``name'', ``age''\chgred{) and} string values \chgred{(e.g.,} ``Alice'', ``Bob'')\chgred{.}
\chgred{The} \emph{symbol table} \chgred{maps} each \chgred{distinct} node label \chgred{bijectively} to a symbol in a unified alphabet $\Sigma$ (Figure~\ref{fig:xbw_construction}(I)).
\chgred{All node labels in $MT$ are converted to symbols in $\Sigma$ (Figure~\ref{fig:xbw_construction}(II)); replacing potentially long string labels with compact symbols makes lexicographic comparison efficient.}
Let $MT'$ denote the \chgred{resulting} merged tree with normalized labels \chgred{and $|MT'|$ nodes}, where \chgb{\chgred{the children of} object \chgred{nodes} are sorted lexicographically, while array \chgred{order} is handled during query processing}.

The jXBW representation of $MT'$ consists of four synchronized arrays \chgred{over the nodes} of $MT'$ (Figure~\ref{fig:xbw_construction}(IV)\chgred{); $A_{\mathit{ids}}$ is stored only for leaves (see below}):

\begin{itemize}
  \item $A_{\mathit{label}}[i]$: the label of the $i$-th node.
  \item $A_{\mathit{last}}[i]$: \chgred{$1$} if the $i$-th node is the rightmost child of its parent, and $0$ otherwise\chgred{; for} the root, $A_{\mathit{last}}[1] = 1$.
  \item $A_{\mathit{leaf}}[i]$: \chgred{$1$} if the $i$-th node is a leaf, and $0$ otherwise.
  \item $A_{\mathit{ids}}[i]$: the \chgred{identifier} set $\mathrm{ids}$ \chgred{of} the $i$-th node.
\end{itemize}

The defining feature of jXBW is the \emph{order} of the entries: nodes are arranged in the lexicographic order of their ancestor label sequences --- the sequence of node labels on the upward path from each node's parent to the root, denoted $A_{\mathit{anc}}[i]$, with $A_{\mathit{anc}}[1] = \varepsilon$ (empty string) for the root. The auxiliary array $A_{\mathit{anc}}$ only defines this order; it is not \chgred{stored} in the final representation.

\label{sec:rank_select}\chgred{\textbf{Rank and select support.} For fast queries, the binary arrays $A_{\mathit{last}}$ and $A_{\mathit{leaf}}$ are indexed by rank and select dictionaries~\cite{jacobson1989space,navarro2016compact,gog2014theory}: for a binary array $B$, $\text{rank}_c(B,i)$ returns the number of occurrences of $c \in \{0,1\}$ in $B[1,i]$, the first $i$ elements of $B$, and $\text{select}_c(B,i)$ returns the position of the $i$-th occurrence of $c$; both take $O(1)$ time using $|B| + o(|B|)$ bits, where the $o(|B|)$ bits hold the auxiliary index. The dictionary is built in $O(|B|)$ time. In practice, the auxiliary index adds about $25\%$ to $37.5\%$ of the input size, depending on the implementation~\cite{gog2014theory}.}

\chgred{The array $A_{\mathit{label}}$ is indexed by a wavelet matrix~\cite{claude2012wavelet}, an alternative to the wavelet tree~\cite{grossi2003high,ferragina2009myriad,navarro2012wavelet} with the same theoretical complexity and a more cache-friendly, practical layout. For an array $A$ over an alphabet of size $\sigma$, both structures support $\text{rank}_c(A,i)$ and $\text{select}_c(A,i)$ for any symbol $c$, defined as above, and $\text{access}(A,i)$, which returns $A[i]$. Each operation takes $O(\log \sigma)$ time, and both structures use $|A|H_0(A) + o(|A|\log\sigma)$ bits, where $H_0(A)$ is the zero-order empirical entropy of $A$. Worked examples are given in \fv{Appendix~\ref{sec:wavelet_details}}{the extended version~\cite{tabei2026jxbwfull}}.}

To \chgred{save} memory, $A_{\mathit{ids}}$ \chgred{stores entries} only \chgred{for} leaf nodes, consecutively. \chgred{The identifier set of the} leaf at position $i$ \chgred{is} $A_{\mathit{ids}}[\text{rank}_1(A_{\mathit{leaf}}, i)]$, where the rank operation maps the leaf position to \chgred{its} index in the compacted array.

\chgred{\textbf{Space.} The arrays $A_{\mathit{label}}$, $A_{\mathit{last}}$, and $A_{\mathit{leaf}}$, together with their wavelet-matrix and rank/select indexes, occupy $O(|MT'| \log \sigma)$ bits; Section~\ref{sec:substructure_search_json} gives the full bound.} While this may be comparable to the pointer-based representation $O(|MT'| \log |MT'|)$ in terms of space complexity for JSONL \chgred{collections} with large vocabularies, jXBW provides significant advantages in query performance.

\chgred{Sorting by $A_{\mathit{anc}}$} yields \chgred{two properties} on which all jXBW operations rely: (i) siblings are stored \chgred{contiguously,} in lexicographic order of their labels; (ii) the $A_{\mathit{last}}$ \chgred{entry of} the rightmost child is always $1$, so each sibling block is a sequence of 0s terminated by a $1$.

For example, in Figure~\ref{fig:xbw_construction}, the \chgred{children} of \chgred{the} nodes labeled \chgred{`A'} (root), \chgred{`A'} (internal), \chgred{`B'}, \chgred{`C'}, \chgred{`F'}, \chgred{`G'}, \chgred{and `I' occupy positions} [2,3], [4,5], [6], [7,8], [9], [10], \chgred{and} [11,12], respectively.
\chgred{The rightmost children, at} positions 3, 5, 6, 8, 9, 10, \chgred{and} 12, \chgred{have} $1$ in $A_{\mathit{last}}$.

\chgred{We construct} jXBW from $MT$ by building the symbol table, extracting the arrays in a depth-first traversal of $MT'$, and stably sorting them by $A_{\mathit{anc}}$ (Figure~\ref{fig:xbw_construction}(I)--(IV));
\fv{Appendix~\ref{sec:jxbw_construction} details this process}{the construction process is detailed in \chgb{the extended version~\cite{tabei2026jxbwfull}}}.

\subsection{Operations on jXBW}
\label{sec:jxbw_ops}
jXBW supports six operations on $MT'$:
\begin{itemize}
\item \textbf{Children$(i)$}: Returns the range $[l, r]$ of \chgred{the} positions \chgred{of} all children of the node at position $i$;
\item \textbf{RankedChild$(i, k)$}: Returns the position \chgred{of} the $k$-th child of the node at position $i$ in lexicographic order;
\item \textbf{CharRankedChild$(i, c, k)$}: Returns the position of the $k$-th child of the node at position $i$ that has label $c$; 
\item \textbf{Parent$(i)$}: Returns the position of the parent of the node at position $i$; 
\item \textbf{TreeIDs$(i)$}: Returns the \chgred{identifier} set of \chgred{the} leaf node at position $i$;
\item \textbf{SubPathSearch$(P)$}: Returns the range $[l, r]$ of \chgred{the} positions \chgred{of all} nodes reachable through path $P = \langle p_1,p_2,\ldots,p_k \rangle$\chgred{.}
\end{itemize}
\chgred{Algorithms}, implementation details, and worked examples for each operation are \chgred{given} in \fv{Appendix~\ref{sec:xbw_operations}}{\chgb{the extended version~\cite{tabei2026jxbwfull}}}.

\chg{These operations enable efficient navigation through the merged tree: \textbf{Children} returns the position range of a node's children; \textbf{RankedChild} and \textbf{CharRankedChild} return the $k$-th child, overall or among the children with a given label, respectively; \textbf{Parent} moves upward; and \textbf{TreeIDs} returns the identifiers of a leaf via $A_{ids}[\text{rank}_1(A_{leaf}, i)]$ on the gap-free, leaf-only array $A_{ids}$.}

The navigation operations \textbf{Children}, \textbf{RankedChild}, \chgred{and} \textbf{Parent} \chgred{take} $O(\log \sigma)$ \chgred{time;}
\textbf{TreeIDs} \chgred{takes} $O(1)$ time using the rank and select dictionary on $A_{leaf}$.

\textbf{SubPathSearch} \chgred{builds on} these \chgred{operations} to find \chgred{the} positions of \chgred{all} nodes reachable through a given \chgred{path $P$}. For example, for $P = \langle$\chgred{`A'}, \chgred{`I'}$\rangle$, \textbf{SubPathSearch$(P)$} returns $[11, 12]$.
A key advantage of \textbf{SubPathSearch} is that it can find all paths \chgred{matching $P$ that start at} any node\chgred{,} not only the root, in $O(|P| \log \sigma )$ time.

\section{Substructure Search Algorithm on jXBW}
\label{sec:substructure_search_json}

\begin{algorithm}[t]
\footnotesize
  \caption{SubstructureSearch algorithm on jXBW}
  \label{alg:subtree_search_adaptive}
  \begin{algorithmic}[1]
    \Function{SubstructureSearch}{$jXBW, Q$}
    \Statex \Comment{\textbf{Step 1: Path Decomposition and Path Matching}}
    \State $paths \gets$ all root-to-leaf paths in $Q$
    \State $ranges \gets \emptyset$
    \For{$P \in paths$}
      \State $[first, last] \gets \Call{SubPathSearch}{P}$ 
      \State $ranges \gets ranges \cup [first, last]$
    \EndFor
    \Statex \Comment{\textbf{Step 2: Finding \chgred{Candidate Roots}}}
    \State $ancestor\_sets \gets \emptyset$
    \For{$i = 1$ to $|paths|$}
      \State $[first, last] \gets ranges[i]$
      \State $ancestor\_sets \gets ancestor\_sets \cup \Call{CompAncestors}{[first, last], paths[i]}$
    \EndFor
    \State $root\_positions \gets \bigcap_{ancestor \in ancestor\_sets} ancestor$
    \Statex \Comment{\textbf{Step 3: \chgred{Collecting Tree Identifiers Adaptively}}}
    \State $all\_matching\_trees \gets \emptyset$
    \For{$root\_pos \in root\_positions$}
      \State $id\_sets \gets$ \\ \hspace*{1.2em}$\Call{CollectPathMatchingIDs}{root\_pos, paths}$
      \State $tree\_ids\_for\_this\_root \gets \bigcap_{ids \in id\_sets} ids$
      \If{$Q$ contains array nodes}
        \Comment{Enforce array order and multiplicity}
        \State $tree\_ids\_for\_this\_root \gets \Call{Verify}{tree\_ids\_for\_this\_root, Q}$
      \EndIf
      \State $all\_matching\_trees \gets all\_matching\_trees \cup tree\_ids\_for\_this\_root$
    \EndFor
    \State \Return $all\_matching\_trees$
  \EndFunction
  \end{algorithmic}
\end{algorithm}

Building on the jXBW operations of the previous section, our substructure search algorithm (Algorithm~\ref{alg:subtree_search_adaptive}) reduces \chgred{tree} matching to path matching and runs in three steps.
First, it decomposes the query tree $Q$ into its root-to-leaf paths and, for each path $P$, applies \textbf{SubPathSearch} to locate the contiguous range of positions whose nodes are reachable through $P$ (Step 1).
Second, it walks up $|P|-1$ levels from every matched position using \textbf{Parent} operations; a position reached from \emph{every} query path can \chgred{be the root of an occurrence of} $Q$, so intersecting the per-path ancestor sets yields the candidate roots (Step 2).
Third, for each candidate root it collects the identifiers of the trees that actually contain $Q$ at that root, \chgred{and verifies array order only when $Q$ contains arrays} (Step 3).
The algorithm thus \chgred{avoids both traversing the whole merged tree and comparing each node's children with those of $Q$ one by one, the two costs of the merged-tree search in Section~\ref{sec:substructure_search_mt}}.\fv{}{ Pseudo-code for the subroutines used below is provided in \chgb{the extended version~\cite{tabei2026jxbwfull}}.}

\begin{figure}[t]
  \centering
  \includegraphics[width=0.92\linewidth]{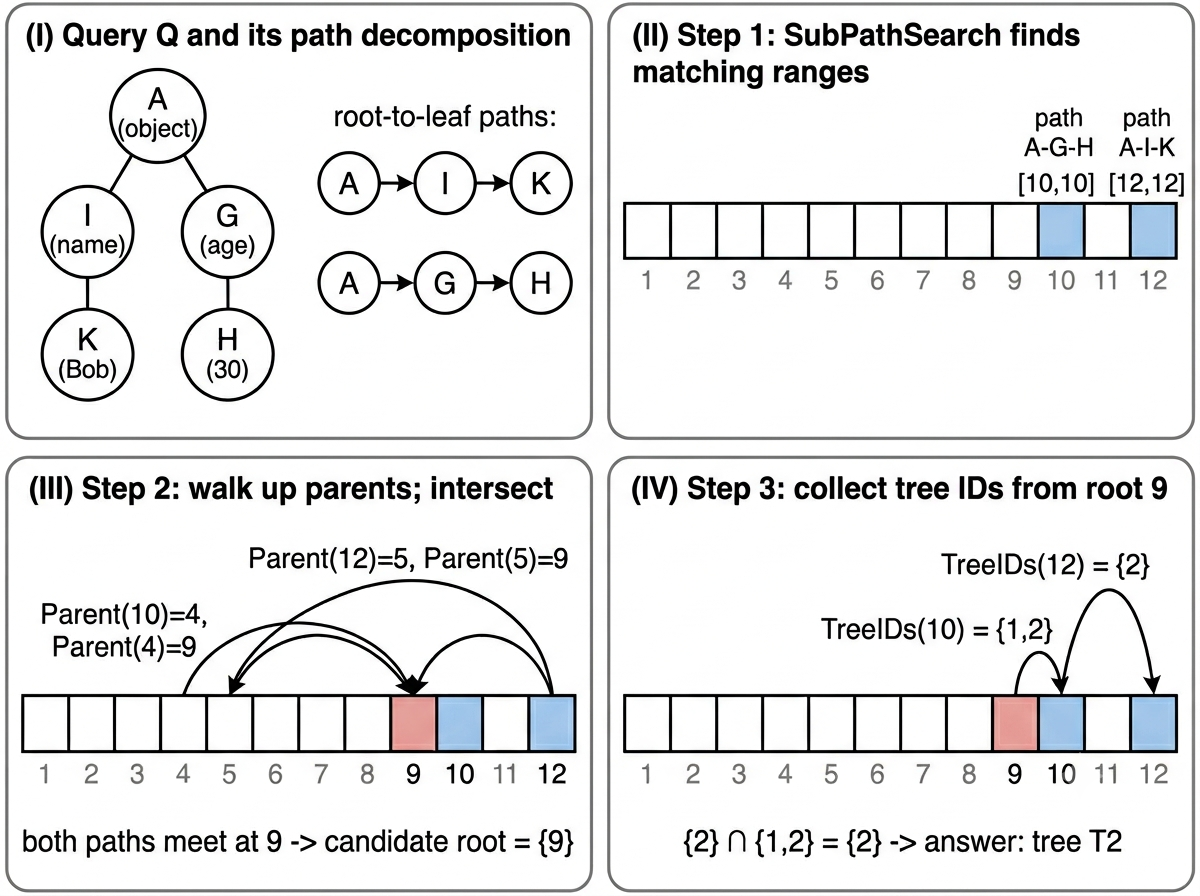}
  \vspace{-0.2cm}
  \caption{Running example of Algorithm~\ref{alg:subtree_search_adaptive} for the query $Q$ of Figure~\ref{fig:query} on the jXBW arrays of Figure~\ref{fig:xbw_construction}: (I) $Q$ is decomposed into root-to-leaf paths; (II) \textbf{SubPathSearch} locates the matching position ranges; (III) walking up $|P|-1$ \chgred{levels} from each matched position and intersecting the ancestor sets yields the candidate root $\{9\}$; (IV) tree identifiers collected from root $9$ give the answer $\{2\}$.}
  \label{fig:substruct_example}
\end{figure}

We now describe the three steps in detail, using the query of Figure~\ref{fig:query} as a running example (Figure~\ref{fig:substruct_example}).

\textbf{Step 1: Path Decomposition and Path Matching.} The algorithm begins by decomposing the query tree $Q$ into all possible root-to-leaf paths $paths$. This decomposition transforms the complex tree matching problem into a collection of simpler path matching problems. For each path $P \in paths$, the algorithm applies the \textbf{SubPathSearch} operation\fvonly{ (Algorithm~\ref{alg:subpath_search})} to identify the range $[first, last]$ \chgred{of positions of} the leaf nodes reachable through $P$. 
This process obtains a set of ranges $ranges$, where each range is computed from a path in $paths$. 

\chgr{For the query $Q$ of Figure~\ref{fig:query}, path decomposition yields $paths = \{\langle object, "name", "Bob" \rangle,$ $\langle object, "age", 30 \rangle\} = \{\langle A, I, K \rangle,$ $\langle A, G, H \rangle\}$ under the \chgred{symbol table} of Figure~\ref{fig:xbw_construction}, and \textbf{SubPathSearch} returns $ranges = \{[12,12],[10,10]\}$.}

\textbf{Step 2: Finding \chgred{Candidate Roots}.} \chgred{If $Q$ occurs at a position $v$, then every root-to-leaf path $P$ of $Q$ leads from $v$ down to a position matched in Step~1, exactly $|P|-1$ levels below $v$; hence $v$ is reached by walking up $|P|-1$ levels from a matched position of \emph{every} path, and the candidate roots are exactly the intersection of the per-path ancestor sets.} For each range $[first, last] \in ranges$ computed from path $P \in paths$, the algorithm performs \textbf{CompAncestors}\fvonly{ (Algorithm~\ref{alg:compute_ancestors})} to trace back $|P| - 1$ levels using \textbf{Parent} operations starting from each position in the range, collecting all ancestor positions that are exactly $|P| - 1$ levels above the leaf positions. All ancestor sets from different paths are collected in $ancestor\_sets$. Finally, $root\_positions$ contains only those positions that are reachable by all query paths, computed as the intersection of all ancestor sets: $root\_positions = \bigcap_{ancestor \in ancestor\_sets} ancestor$.

\chgr{For $ranges = \{[12,12],[10,10]\}$, \chgred{walking} up $|P| - 1 = 2$ levels gives $\textbf{Parent}(12) = 5$, $\textbf{Parent}(5) = 9$ and $\textbf{Parent}(10) = 4$, $\textbf{Parent}(4) = 9$, so $ancestor\_sets = \{\{9\}, \{9\}\}$ and $root\_positions = \{9\}$.}

\textbf{Step 3: \chgred{Collecting Tree Identifiers Adaptively}.} For each $root\_pos \in root\_positions$, \textbf{CollectPathMatchingIDs}\fvonly{ (Algorithm~\ref{alg:collect_path_matching_ids})} navigates \chgred{the} query paths \chgred{from $root\_pos$} using \textbf{CharRankedChild} \chgred{and collects tree identifiers only from the leaves reached. Intersecting these identifier sets keeps only the trees in which $Q$ occurs at that root}. \chgb{\chgred{Path} decomposition does not constrain the \chgred{order of} array elements, \chgred{so} for queries containing arrays\chgred{, an array verification step checks} the candidate tree identifiers against the original JSONL \chgred{collection} to enforce array ordering and multiplicity; array-free queries \chgred{skip this step}.} The final result aggregates tree identifiers from all \chgred{candidate} roots using set union: $all\_matching\_trees = all\_matching\_trees \cup tree\_ids\_for\_this\_root$.

\chgr{For $root\_positions = \{9\}$, \textbf{CollectPathMatchingIDs} navigates each path from position $9$: path $\langle A,I,K \rangle$ reaches leaf position $12$ with $\textbf{TreeIDs}(12) = \{2\}$, and path $\langle A,G,H \rangle$ reaches leaf position $10$ with $\textbf{TreeIDs}(10) = \{1,2\}$. Their intersection \chgred{gives the answer} $all\_matching\_trees = \{2\}$.}

\chg{\textbf{Time Complexity Analysis.} Let $p$ denote the number of root-to-leaf paths in \chgred{the} query tree $Q$, $d$ the average path depth, $\sigma$ the alphabet size, $r$ the total number of matching leaf positions across all paths, and $c$ the number of \chgred{candidate roots}. Step 1 performs $p$ \textbf{SubPathSearch} operations in $O(p \cdot d \cdot \log \sigma)$ total time; Step 2 executes $r \cdot d$ \textbf{Parent} operations in $O(r \cdot d \cdot \log \sigma)$ time; Step 3 takes $O(c \cdot p \cdot d \cdot \log \sigma)$ time via \textbf{CollectPathMatchingIDs}. \chgb{Step 3 additionally intersects the \chgred{identifier} sets attached to the matched positions and enumerates the result; let $I_{\mathit{ids}}$ denote the total number of identifier entries examined during these intersections and output collection.} The overall complexity is therefore $O((p + r) \cdot d \cdot \log \sigma + c \cdot p \cdot d \cdot \log \sigma \chgb{{} + I_{\mathit{ids}} + V})$ (Table~\ref{tab:complexity}), \chgb{where \chgred{$V$ is the total verification cost for queries containing arrays and $V = 0$ for array-free queries}. The bound \chgred{has no term for scanning} the collection: it is candidate- and output-sensitive rather than a function of the collection size $N$}.}

\chg{\textbf{Space Complexity Analysis.} jXBW occupies $O(|MT'| \log \sigma + L + |A_{\mathit{ids}}| \log N)$ space, where $L$ is the total size of the distinct JSON labels in the \chgred{symbol table} and $|A_{\mathit{ids}}|$ is the total number of stored tree identifiers; in practice, the \chgred{symbol table} dominates memory \chgred{use} (Section~\ref{sec:results}).}

\section{Experiments}
\label{sec:experiments}

\begin{table*}[ht]
\centering
    \caption{Summary of the datasets. \chgred{\#Distinct Keys: number of different key names in the dataset.}}
    \begin{tabular}{lrrrrr}
\toprule
    \textbf{Dataset} & \textbf{File Size} & \textbf{\#Objects} & \textbf{\chgred{\#Distinct Keys}} & \textbf{Avg. Tree} & \textbf{Array Queries} \\
    & \textbf{(MB)} & & & \textbf{Depth} & \textbf{(\%)} \\
    \midrule
    movies & 22 & 36,273 & 9 & 2.99 & 0.0 \\
    electric\_vehicle\_population & 199 & 247,344 & 28 & 2.00 & 0.0 \\
    border\_crossing\_entry & 104 & 401,566 & 1 & 2.00 & 100.0 \\
    mta\_nyct\_paratransit & 242 & 1,572,461 & 1 & 2.00 & 100.0 \\
    osm\_data\_newyork & 318 & 1,090,245 & 2,001 & 2.47 & 0.0 \\
    osm\_data\_tokyo & 1,100 & 6,602,928 & 2,496 & 2.36 & 0.0 \\
    pubchem & 7,066 & 1,000,000 & 53 & 6.00 & 0.0 \\
    \bottomrule
\end{tabular}
\label{tab:jsonl_analysis}
\end{table*}

  \begin{table*}[htbp]
\centering
    \caption{Average substructure search time per query (ms). \emph{Avg Hits}\chgred{:} average number of matching objects per query. \chgred{Times} are end-to-end (matching + \chgred{tree} identifier collection)\chgred{. The fastest time in each row is in bold}.}
\label{tab:search_performance}
\label{tab:query-time}
    \begin{tabular}{lrrrrrr}
\toprule
    \textbf{Dataset} & \textbf{Avg Hits} & \textbf{jXBW} & \textbf{Ptree} & \textbf{SucTree} & \textbf{Saxon} & \chgb{\textbf{DuckDB}} \\
\midrule
    movies & 70.7 & \chgred{\textbf{0.016 $\pm$ 0.020}} & 0.265 $\pm$ 0.025 & 2.054 $\pm$ 0.083 & 3.958 $\pm$ 0.962 & 17.041 $\pm$ 0.458 \\
    electric\_vehicle\_population & 418.6 & \chgred{\textbf{0.038 $\pm$ 0.014}} & 1.278 $\pm$ 0.037 & 10.264 $\pm$ 0.276 & 40.224 $\pm$ 7.729 & 130.786 $\pm$ 3.844 \\
    border\_crossing\_entry & 1.0 & \chgb{\chgred{\textbf{0.015 $\pm$ 0.004}}$^{2}$} & 1.407 $\pm$ 0.035 & 13.664 $\pm$ 0.579 & 49.958 $\pm$ 6.235 & 132.962 $\pm$ 1.718 \\
    mta\_nyct\_paratransit & 1.0 & \chgb{\chgred{\textbf{0.012 $\pm$ 0.004}}$^{2}$} & 5.619 $\pm$ 0.173 & 48.716 $\pm$ 1.447 & 167.929 $\pm$ 23.793 & 322.165 $\pm$ 3.036 \\
    osm\_data\_newyork & 16.4 & \chgred{\textbf{0.022 $\pm$ 0.009}} & 12.464 $\pm$ 0.421 & 97.739 $\pm$ 2.553 & 269.580 $\pm$ 18.400 & 343.559 $\pm$ 8.272 \\
    osm\_data\_tokyo & 26.8 & \chgred{\textbf{0.026 $\pm$ 0.012}} & 73.127 $\pm$ 1.246 & 373.371 $\pm$ 6.556 & 1,954.94 $\pm$ 72.68 & 1,151.945 $\pm$ 21.929 \\
    pubchem & 268.6 & \chgred{\textbf{0.043 $\pm$ 0.018}} & 26.453 $\pm$ 0.692 & 205.189 $\pm$ 6.943 & N/A$^{1}$ & 6,412.967 $\pm$ 17.375 \\
\bottomrule
\end{tabular}
\par\smallskip
\footnotesize $^{1}$ \chgb{Saxon did not complete the 1{,}000-query workload within 24~hours.} \chgb{$^{2}$ For the two array-query datasets, \chgred{search time} includes exact array verification (Section~\ref{sec:substructure_search_json}).}
\end{table*}

\begin{table*}[htbp]
\centering
\caption{Index memory usage (MB) and construction time (s). \chgred{For} the tree-based methods\chgred{, memory} includes the shared \chgred{symbol table, and} construction time includes building the individual tree structures and merging them; Saxon times \chgred{cover} XML parsing and its native index construction\chgb{; DuckDB times cover JSONL parsing, loading, and checkpointing}.}
\label{tab:memory_usage_with_saxon}
\label{tab:index-memory}
\label{tab:construction_time}
\label{tab:construction-time}
\begin{tabular}{l|rrrrr|rrrrr}
\toprule
 & \multicolumn{5}{c|}{\textbf{Memory (MB)}} & \multicolumn{5}{c}{\textbf{Construction time (s)}} \\
\textbf{Dataset} & \textbf{jXBW} & \textbf{Ptree} & \textbf{SucTree} & \textbf{Saxon} & \chgb{\textbf{DuckDB}} & \textbf{jXBW} & \textbf{Ptree} & \textbf{SucTree} & \textbf{Saxon} & \chgb{\textbf{DuckDB}} \\
\midrule
movies & 56 & 66 & 50 & 329 & 12 & 1.845 & 1.821 & 1.973 & 0.787 & 0.090 \\
electric\_vehicle\_population & 172 & 219 & 147 & 1,664 & 81 & 159.066 & 158.335 & 159.071 & 9.599 & 0.480 \\
border\_crossing\_entry & 206 & 255 & 182 & 1,593 & 45 & 358.634 & 359.722 & 362.992 & 13.169 & 0.343 \\
mta\_nyct\_paratransit & 691 & 875 & 594 & 3,241 & 92 & 4,835.966 & 5,059.798 & 5,053.181 & 55.093 & 0.855 \\
osm\_data\_newyork & 842 & 1,153 & 624 & 4,180 & 306 & 1,758.649 & 1,759.208 & 1,759.115 & 63.900 & 0.609 \\
osm\_data\_tokyo & 2,964 & 4,264 & 2,072 & 8,495 & 926 & 47,273.986 & 46,668.465 & 46,496.468 & 90.073 & 2.485 \\
pubchem & 3,880 & 4,702 & 3,453 & 64,840 & 7,173 & 16,097.010 & 16,320.957 & 15,856.235 & 629.764 & 7.214 \\
\bottomrule
\end{tabular}
\par\smallskip
\footnotesize \chgb{For \chgred{the two array-query} datasets, jXBW's \chgred{array} verification \chgred{also} keeps the original JSONL \chgred{collection} in memory (border\_crossing\_entry: 158~MB, mta\_nyct\_paratransit: 334~MB), \chgred{which is excluded from} the index sizes above.}
\end{table*}

\subsection{Setup}

In this section, we demonstrate the effectiveness of jXBW's substructure search with large datasets. We used seven datasets, as shown in Table~\ref{tab:jsonl_analysis}. These datasets were obtained from various publicly accessible sources and converted to JSONL format for our experiments. The datasets span diverse domains including entertainment, transportation, government services, geographical information, and chemical compounds, providing a comprehensive evaluation of our jXBW approach across different data characteristics and structural complexities. 

\chg{The seven datasets comprise movie metadata scraped from Wikipedia~\cite{wikipedia_movie_data}, electric-vehicle registrations~\cite{electric_vehicle_data} and border-crossing records~\cite{border_crossing_data} from Kaggle, paratransit service records from the U.S. Government data catalog~\cite{paratransit_data}, OpenStreetMap \chgred{geospatial} objects for New York and Tokyo~\cite{osm_data}, and chemical compound records from PubChem~\cite{pubchem_data}; Table~\ref{tab:jsonl_analysis} summarizes their sizes, numbers of objects \chgred{and distinct keys}, and tree depths. For pubchem, \chgred{we converted} the original SDF (Structure Data File) format to JSONL, where each compound is represented as a tree with 53 \chgred{distinct keys} covering structural and molecular properties.}

\chgr{For each dataset, we randomly sampled 1{,}000 \chgred{query trees} of depth 2--4 \chgred{as connected substructures of} the original trees, so that every query has a non-empty result set; the \chgred{query trees are derived from} both JSON \chgred{objects} (starting with `\{`) and JSON \chgred{arrays} (starting with `[`).}

\chgr{We compared jXBW with the following methods; \chgred{Table~\ref{tab:complexity} summarizes} their index strategies and theoretical complexities.}
\textbf{Ptree} is a merged tree representation using pointers. 
The JSONL \chgred{collection} is converted to tree structures and merged into a single merged tree as introduced in Section~\ref{sec:merged_tree}, implemented using standard pointer-based data structures.
\textbf{SucTree} is a succinct representation based on the approach by Lee et al.~\cite{lee2020sjson}.
In Lee et al.'s approach, each individual JSON object is represented using a succinct ordinal-tree data structure based on LOUDS~\cite{jacobson1989space}.
We extended this idea to represent the merged tree.
Both Ptree and SucTree perform substructure search by traversing the merged tree structure as described in Section~\ref{sec:substructure_search_mt}.

\textbf{Saxon}~\cite{saxon_processor} is a high-performance XQuery processor widely used in enterprise applications, providing a mature implementation of XQuery 3.1 and XPath 3.1 standards. We used Saxon-HE (Home Edition), the open-source version of Saxon, which provides full XQuery and XSLT processing capabilities. For Saxon-based comparison, we first converted each \chgred{JSON} object to XML format, then performed equivalent substructure searches using XQuery expressions that match the \chgred{query trees} used in our jXBW queries.

\chgb{\textbf{DuckDB}~\cite{raasveldt2019duckdb} (v1.5.5) stores each JSONL file in a JSON column. Each query uses \texttt{json\_contains} for structural containment --- an unindexed recursive check whose worst-case cost is proportional to the product of \chgred{object} and query sizes. For arrays with two or more elements, we additionally use a \texttt{list\_reduce} predicate to perform a greedy, duplicate-sensitive, order-preserving subsequence match. \chgred{DuckDB thus performs all} matching inside \chgred{its} SQL engine, without client-side post-processing.}

We implemented jXBW, Ptree, and SucTree in C++.
We used the C++ \chgred{SDSL-lite} library~\cite{gog2014theory} for rank and select \chgred{dictionaries} in jXBW and SucTree and for wavelet \chgred{trees} in jXBW.
\chgr{All experiments ran single-threaded on a single CPU core of an Apple M4 Max (128~GB unified memory); \chgb{\chgred{we stopped} a method if it did not complete a dataset's full 1{,}000-query workload within 24~hours}.}

\chgred{We report all} results as mean \(\pm\) standard deviation over 1{,}000 queries per dataset.
All sampling used a fixed random seed (\texttt{seed=42}) for reproducibility. 
We report time in milliseconds (ms) and seconds (s), and size in megabytes (MB) unless otherwise noted.

\subsection{Experimental Results}
\label{sec:results}
\chg{We evaluate jXBW in terms of query performance, memory efficiency, and construction time.}

\textbf{Query Performance:} Table~\ref{tab:search_performance} presents the substructure search performance comparison, where jXBW measurements represent complete end-to-end search time. 
As shown in Table~\ref{tab:jsonl_analysis}, border\_crossing\_entry and mta\_nyct\_paratransit consist entirely of queries containing \chgred{arrays}, while the other datasets contain no array queries.
\chg{jXBW delivers the fastest query times on all datasets, and the advantage \chgred{tends to grow} with \chgred{the number of objects. On} pubchem (1M compounds), jXBW answers queries in the sub-millisecond range, with speedups of over 615$\times$ versus Ptree, 4,772$\times$ versus SucTree, \chgred{and} over 2$\times$10$^6$ versus Saxon, \chgb{which did not complete the 1{,}000-query workload within 24~hours}\chgb{. Across all datasets\chgred{,} jXBW is $1.1\times10^{3}$--$1.5\times10^{5}$ times faster than DuckDB, \chgred{with the largest speedup} on pubchem}. \chgb{Even on the two \chgred{array-query} datasets (border\_crossing\_entry, mta\_nyct\_paratransit), \chgred{jXBW, including exact array verification, achieves} $94\times$/$468\times$ speedups over Ptree and $911\times$/$4{,}060\times$ over SucTree.}}

\chgb{jXBW remains fast even for queries with hundreds of matching objects (e.g., \texttt{pubchem}: 268.6 \emph{Avg Hits}; Table~\ref{tab:search_performance}), while avoiding a full scan of the collection.}

\chg{\textbf{Memory Efficiency:} \chgred{For} the tree-based methods, the shared symbol table \chgred{dominates index memory} (e.g., $2{,}758$ of jXBW's $3{,}880$~MB on pubchem\chgred{; Table~\ref{tab:memory_usage_with_saxon}}). SucTree is the most compact \chgred{tree-based method}, \chgred{and} jXBW stays well below both Ptree and Saxon (e.g., $2{,}964$ \chgred{versus} $4{,}264$ and $8{,}495$~MB on osm\_data\_tokyo) while supporting much faster queries. \chgb{DuckDB is the most compact \chgred{overall} except on pubchem, where jXBW uses $3{,}880$ versus $7{,}173$~MB.}}

\chg{\textbf{Construction Time:} jXBW's construction time is competitive with the other tree-based \chgred{methods} because merged tree construction dominates the total (e.g., $47{,}211$ of jXBW's $47{,}274$\chgred{~s} on osm\_data\_tokyo\chgred{; Table~\ref{tab:construction_time}}); Saxon builds faster owing to its optimized XML parsing pipeline\chgb{, and DuckDB fastest by a wide margin ($2.5$ versus $47{,}274$~s on osm\_data\_tokyo). jXBW thus trades \chgred{a} one-time construction cost for orders-of-magnitude \chgred{faster queries}}.}

\subsection{Case Study: Chemical Compound Analysis with GPT}
\label{sec:case_study}
\begin{figure}[t]
  \centering
  \includegraphics[width=0.60\linewidth]{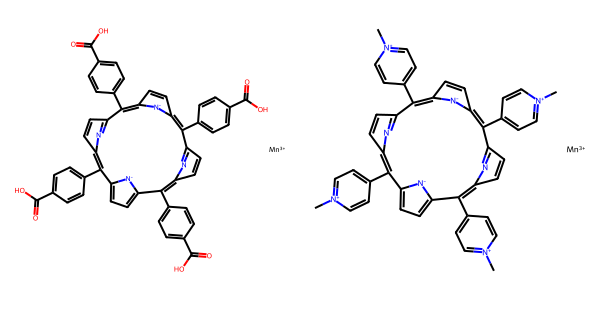}
  \vspace{-0.5cm}
  \caption{Representative molecular structures retrieved by jXBW with the query $\{\text{"structure"}: \{\text{"atoms"}: [\{\text{"symbol"}: \text{"N"}, \text{"charge"}: 1\}]\}\}$. Both exhibit cationic nitrogen centers (N$^+$), extended conjugation, and potential redox activity.}
  \label{fig:chemical_examples}
\end{figure}

To evaluate jXBW-based substructure search in a real-world scientific workflow, we queried the \chgred{pubchem} dataset (Table~\ref{tab:jsonl_analysis}) for compounds containing cationic nitrogen atoms (N$^+$) with the \chgred{query tree} $\{\text{"structure"}: \{\text{"atoms"}: [\{\text{"symbol"}: \text{"N"}, \text{"charge"}: 1\}]\}\}$.
The query retrieved 838 compounds\chgred{; we passed} 10 randomly sampled \chgred{ones} to GPT-4. GPT-4 identified recurring structural motifs (quaternary and pyridinium nitrogen centers, porphyrin-like aromatic macrocycles, transition-metal coordination) and hypothesized pharmacological applications including photosensitization in photodynamic therapy and DNA/RNA binding. Figure~\ref{fig:chemical_examples} shows two representative hits: a Mn$^{3+}$-coordinated porphyrin derivative with carboxyphenyl groups (left) and a pyridinium-substituted porphyrin complex (right). This \chgred{case study} illustrates how fast structural retrieval can \chgred{help an LLM triage compounds and generate structure--function hypotheses}; the detailed analysis \chgred{appears} in \fv{Appendix~\ref{sec:case_study_details}}{\chgb{the extended version~\cite{tabei2026jxbwfull}}}.

\section{Conclusion and Future Work}

\chg{We presented jXBW, a compressed index for substructure search over large-scale JSONL collections, built from a merged tree representation, an XBW-based succinct encoding, and a three-phase search algorithm. jXBW has the following appealing properties:
\begin{enumerate}
  \item \textbf{Scalability:} jXBW handles collections of up to $6.6 \times 10^6$ objects, with a one-time construction cost that amortizes over typical workloads (Section~\ref{sec:results}).
  \item \textbf{Query-dependent complexity:} \chgb{query time is candidate- and output-sensitive and avoids a linear scan over the collection}, yielding sub-millisecond queries and speedups growing from $16\times$ on the smallest dataset to $2{,}800\times$ on the largest (Sections~\ref{sec:substructure_search_json} and~\ref{sec:results}).
  \item \textbf{\chgred{Space efficiency}:} \chgb{the main search operations run on the compressed representation}, with memory below pointer-based indexes and competitive with succinct ones (Section~\ref{sec:results}).
  \item \chgb{\textbf{Generality:} a single matcher handles both ordered and unordered semantics across diverse \chgred{collections} --- from chemistry to geospatial data in our seven datasets --- and supports LLM-driven analysis, as demonstrated in the GPT case study (Section~\ref{sec:case_study}).}
\end{enumerate}}

\chg{The success of jXBW opens several promising directions for future work: extending path decomposition to support \chgred{approximate substructure matching based on edit distance}, leveraging advances in approximate string matching on compressed indices; applying the merged tree representation to other semi-structured formats such as Protocol Buffers, Apache Avro, and MessagePack; developing distributed versions of jXBW using consistent hashing and parallel query processing to handle extremely large \chgred{collections} (e.g., $>$100~GB); and combining structural matching with semantic similarity measures to enable LLM-guided query expansion and domain-aware retrieval.}

\chgr{Finally, the integration of jXBW with LLMs demonstrated in our case study suggests that combining efficient structured-data retrieval with neural reasoning can be a key enabler for symbolic--neural hybrid systems.}

\fv{}{\balance}
\bibliographystyle{IEEEtran}
\bibliography{biblio}

\iffullversion

\newpage

\appendices

\section{JSON Syntax in Backus--Naur Form}
\label{sec:json_bnf}

\chgred{The following Backus--Naur Form (BNF) grammar formally defines each JSON object $O_i$:}

\begin{minipage}{0.95\linewidth}
\begin{ttfamily}
<object>  ::= \{ <members> ? \} \\
<members> ::= <pair> ( , <pair> )* \\
<pair>    ::= <string> : <value>
\end{ttfamily}
\end{minipage}

Here, \texttt{<value>} is one of \texttt{<string>}, \texttt{<number>}, \texttt{<object>}, \texttt{<array>}, \texttt{true}, \texttt{false}, or \texttt{null}.
An \texttt{<object>} may include \texttt{<members>}, which consist of one or more unordered \texttt{<pair>} elements.
Each \texttt{<pair>} \chgred{associates} a \texttt{<string>} \chgred{key with a} value \chgred{of} any valid JSON data type defined above.
An \texttt{<array>} \chgred{is} an ordered sequence of elements of any valid JSON data type.

\section{Tree Merging Algorithm}
\label{sec:merge_trees}
\chgred{Algorithm~\ref{alg:tree_merge} gives the pseudo-code of \textbf{MergeTrees}.}

\begin{algorithm}[h]
  \caption{MergeTrees algorithm for merging two trees $T$ and $T'$.}
  \label{alg:tree_merge}
  \begin{algorithmic}[1]
  \Function{MergeTrees}{$T, T'$}
      \If{$T'$ has no nodes}
          \State \Return $T$
      \EndIf
      
      \If{$T$ has no nodes}
          \State $T \gets \Call{CopyTree}{T'}$ \Comment{Copy entire tree structure}
          \State \Return $T$
      \EndIf
      
      \If{$T.root.label \neq T'.root.label$}
          \State \Call{AddSubtree}{$T.root$, $T'.root$} \Comment{Add the tree $T'$ as a new child of the root of $T$}
      \Else
          \Comment{Same root labels}
          \State \Call{MergeRecursive}{$T.root$, $T'.root$} \Comment{Merge the trees recursively}
      \EndIf
      
      \State \Return $T$
  \EndFunction
  
  \Function{MergeRecursive}{$node, node'$}
      \If{$node'$ is leaf}
          \Comment{Merge tree IDs from leaf node}
          \State $node.ids \gets node.ids \cup node'.ids$
          \State \Return
      \EndIf
      
      \For{each $child'$ in $node'.children$}
          \State $found \gets$ false
          \For{each $child$ in $node.children$}
              \If{$child.label = child'.label$}
                  \State \Call{MergeRecursive}{$child$, $child'$}
                  \State $found \gets$ true
                  \State \textbf{break}
              \EndIf
          \EndFor
          
          \If{$found = false$}
              \Comment{No matching child found}
              \State \Call{AddSubtree}{$node$, $child'$} \Comment{Add the subtree to the node as a new child}
          \EndIf
      \EndFor
  \EndFunction
  \Function{AddSubtree}{$parent, node'$}
      \State Create a new node $new\_node$ with label $node'.label$
      \State $new\_node.ids \gets node'.ids$
      \State $parent.children \gets parent.children \cup \{new\_node\}$
      
      \For{each $child'$ in $node'.children$}
          \State \Call{AddSubtree}{$new\_node$, $child'$}
      \EndFor
  \EndFunction
  \end{algorithmic}
\end{algorithm}

\paragraph{Cost analysis}
\chgred{Let $M_{\mathrm{tot}} = \sum_{i=1}^{N} |T_i|$ be the total number of nodes in $N$ trees $T_1, T_2, \dots, T_N$ that are merged sequentially.} In typical cases\chgred{,} where node labels are distinct or only partially overlapping, each merge takes $O(|T| + |T'|)$ time, \chgred{for} a total cost of $O(M_{\mathrm{tot}} \cdot N)$. \chgred{In the worst case}, where all node labels are identical and \chgred{the} tree structures are highly similar, each merge may \chgred{take} \chgred{$O(|T| \cdot |T'|)$} time \chgred{because it must} examine all possible node pairings, \chgred{so the} total cost can reach $O(M_{\mathrm{tot}}^2)$.

\chgred{{\bf Divide-and-conquer strategy.} To reduce this cost, we merge the trees in a balanced hierarchy.}
At each level of recursion, \chgred{we pair the} trees and \chgred{merge the pairs} in parallel, \chgred{repeating} until a single tree remains.
\chgred{Because} the total number of nodes $M_{\mathrm{tot}}$ \chgred{does not change during} merging and each level processes \chgred{every node} once, \chgred{each} level \chgred{costs} $O(M_{\mathrm{tot}})$.
\chgred{With} $O(\log N)$ \chgred{levels}, the overall merging cost \chgred{drops} to $O(M_{\mathrm{tot}} \log N)$.
\chgred{By avoiding the large intermediate trees that accumulate in sequential merging, this} strategy significantly improves efficiency and ensures better scalability \chgred{for} a large number of input trees.

\section{Merged Tree Substructure Search Algorithms}
\label{sec:mt_algorithms}

This appendix \chgred{details} the algorithms for substructure search on \chgred{the} merged tree \chgred{summarized} in Section~\ref{sec:substructure_search_mt}.

\begin{algorithm}[t]
  \caption{SubstructureSearch algorithm on \chgred{the} merged tree for finding all tree identifiers of trees $T_i$ that contain query tree $Q$ as a \chgred{substructure}.}
  \label{alg:substructure_search_mt}
  \begin{algorithmic}[1]
  \Function{SubstructureSearchMT}{$MT, Q$}
    \State $solutions \gets \emptyset$
    \State $root\_label \gets$ label of root node in $Q$
    \Statex
    
    \Statex \Comment{\textbf{Step 1: Candidate Finding}}
    \State $candidates \gets \Call{FindCandidateNodes}{MT, root\_label}$
    \Statex
    
    \Statex \Comment{\textbf{Step 2: Substructure Matching}}
    \For{each $candidate \in candidates$}
      \State $leaf\_id\_sets \gets \Call{MatchSubtree}{candidate, Q.root}$
      \If{$leaf\_id\_sets \neq \emptyset$}
        \Statex \Comment{\textbf{Step 3: Solution Identification}}
        \State $intersection\_ids \gets \bigcap_{ids \in leaf\_id\_sets} ids$
        \State $solutions \gets solutions \cup intersection\_ids$
      \EndIf
    \EndFor
    
    \State \Return $solutions$
  \EndFunction
  \end{algorithmic}
\end{algorithm}

\begin{algorithm}[t]
  \caption{FindCandidateNodes algorithm for finding all nodes in \chgred{the} merged tree with \chgred{a} specified label.}
  \label{alg:find_candidate_nodes}
  \begin{algorithmic}[1]
  \Function{FindCandidateNodes}{$MT, root\_label$}
    \State $candidates \gets \emptyset$
    \State $\Call{TraverseMT}{MT.root, root\_label, candidates}$
    \State \Return $candidates$
  \EndFunction
  \Statex
  \Function{TraverseMT}{$node, target\_label, candidates$}
    \If{$node.label = target\_label$}
      \State $candidates \gets candidates \cup \{node\}$
    \EndIf
    \For{each $child \in node.children$}
      \State $\Call{TraverseMT}{child, target\_label, candidates}$
    \EndFor
  \EndFunction
  \end{algorithmic}
\end{algorithm}

\begin{algorithm}[t]
  \caption{MatchSubtree algorithm for matching \chgred{the} subtree \chgred{of $MT$ rooted at} $mt\_node$ against \chgred{the subtree of $Q$ rooted at} $q\_node$.}
  \label{alg:match_subtree}
  \begin{algorithmic}[1]
  \Function{MatchSubtree}{$mt\_node, q\_node$}
    \If{both $q\_node$ and $mt\_node$ are leaf}
      \State \Return $\{mt\_node.ids\}$
    \ElsIf{$q\_node$ is leaf and $mt\_node$ is not leaf}
      \State \Return $\emptyset$
    \ElsIf{$q\_node$ is not leaf and $mt\_node$ is leaf}
      \State \Return $\emptyset$
    \EndIf
    
    \Statex \Comment{Both nodes are internal: match children in order}
    \State $all\_leaf\_ids \gets \emptyset$
    \State $mt\_idx \gets 1$, $q\_idx \gets 1$
    \While{$q\_idx \leq |q\_node.children|$ and $mt\_idx \leq |mt\_node.children|$}
      \State $q\_child \gets q\_node.children[q\_idx]$
      \State $mt\_child \gets mt\_node.children[mt\_idx]$
      \If{$mt\_child.label = q\_child.label$}
        \State $result \gets\Call{MatchSubtree}{mt\_child, q\_child}$
        \If{$result \neq \emptyset$}
          \State $all\_leaf\_ids \gets all\_leaf\_ids \cup result$
          \State $q\_idx \gets q\_idx + 1$
        \Else
          \State \Return $\emptyset$
        \EndIf
      \EndIf
      \State $mt\_idx \gets mt\_idx + 1$
    \EndWhile
    \If{$q\_idx \leq |q\_node.children|$}
      \State \Return $\emptyset$ \Comment{Some query children were not matched}
    \EndIf
    
    \State \Return $all\_leaf\_ids$
  \EndFunction
  \end{algorithmic}
\end{algorithm}

\section{Wavelet Tree and Matrix: Details}
\label{sec:wavelet_details}

This appendix expands on the wavelet tree and wavelet matrix summarized in Section~\ref{sec:rank_select}.

The wavelet tree \chgred{reduces} queries on a large alphabet to a series of binary queries, \chgred{each} answered efficiently \chgred{by} rank and select on bit arrays.
Consider \chgred{the} array $A = [a, b, a, c]$ \chgred{over the} alphabet $\{a, b, c\}$.
The wavelet tree splits the alphabet into $\{a\}$ and $\{b, c\}$, creating the bit array $[0, 1, 0, 1]$ (0 \chgred{for $a$;} 1 \chgred{for $b$} or \chgred{$c$}).
To count how many times $a$ appears in the first three positions, we compute $\text{rank}_0([0, 1, 0, 1], 3) = 2$.
\chgred{Counting elements thus becomes counting bits.}
The operation $\text{access}(A, i)$ retrieves the element at position $i$ \chgred{of} $A$ by following a path from the root to a leaf, guided by the bit array at each level.

\chgred{The} wavelet matrix \chgred{instead} stores all bit arrays \chgred{as} a matrix\chgred{, with one row for} each level of the original tree.
This \chgred{layout} eliminates explicit tree navigation and reduces memory overhead, making \chgred{the wavelet matrix} particularly suitable for large-scale applications.

\section{jXBW Construction}
\label{sec:jxbw_construction}

This appendix details the three-step construction of jXBW from the merged tree $MT$, illustrated in Figure~\ref{fig:xbw_construction}.

\textbf{Step 1: Symbol Table Creation and Label Normalization.}
\chgred{We build a} symbol table that bijectively maps each unique node label of $MT$ to a symbol in the unified alphabet $\Sigma$ (Figure~\ref{fig:xbw_construction}(I)) and \chgred{convert} all node labels of $MT$ to $\Sigma$ (Figure~\ref{fig:xbw_construction}(II)). \chgred{We then sort the} children of each object node lexicographically by label \chgred{and keep} the children of array nodes \chgred{in} their original order, \chgred{which yields} the normalized merged tree $MT'$.

\textbf{Step 2: Array Construction via Depth-First Traversal.}
\chgred{A single depth-first traversal of $MT'$ populates the} five synchronized arrays $A_{\mathit{label}}$, $A_{\mathit{anc}}$, $A_{\mathit{last}}$, $A_{\mathit{leaf}}$, and $A_{\mathit{ids}}$ defined in Section~\ref{sec:jxbw_structure}, one entry per visited node (Figure~\ref{fig:xbw_construction}(III)).

\textbf{Step 3: Stable Lexicographic Sorting.}
\chgred{We stably sort all} five arrays \chgred{together} by the lexicographic order of the sequences in $A_{\mathit{anc}}$ (Figure~\ref{fig:xbw_construction}(IV)) and \chgred{then discard} $A_{\mathit{anc}}$. The depth-first extraction order and the stability of the sort guarantee properties (i)--(ii) of Section~\ref{sec:jxbw_structure}: children of the same parent share the same ancestor label sequence and therefore remain contiguous in the lexicographic label order inherited from $MT'$, with each sibling block terminated by its rightmost child marked $1$ in $A_{\mathit{last}}$.

\section{\texorpdfstring{\chgred{jXBW}}{jXBW} Operations: Detailed Algorithms and Examples}
\label{sec:xbw_operations}

This appendix \chgred{gives} algorithms, implementation \chgred{details}, and worked examples for the six \chgred{jXBW} operations introduced in Section~\ref{sec:jxbw_ops}.

\subsection{Children Operation}

\textbf{Children$(i)$}: \chgred{This operation} exploits the XBW property that siblings are \chgred{contiguous} in the arrays. We identify the label at position $i$ (the parent label \chgred{of} the children we seek), locate the first occurrence of nodes having this parent label, and then use the binary encoding in $A_{last}$ to determine the exact range where all children of position $i$ are stored. \chgred{The operation proceeds as follows:}
(i) Compute the \chgred{label} $c$ at position $i$ in $A_{label}$ using the access operation $\text{access}(A_{label}, i)$ on the wavelet matrix.
(ii) \chgred{Compute} \chgred{$y = F(c)$}\chgred{,} the first position such that the parent of the node at $y$ has label $c$. \chgred{To compute it, we construct} an array $A_{pf}$ where $A_{pf}[j]$ equals the first \chgred{label} of $A_{anc}[j]$ if $j \neq 1$ and $A_{pf}[j] = 0$ otherwise\chgred{, and compute} $F(c)$ as $\text{select}_c(A_{pf}, 1)$.
(iii) \chgred{Compute the} children range $[l, r]$\chgred{,} the contiguous block of siblings\chgred{, with rank and select operations}:
$l = \text{select}_1(A_{last}, z + s - 1) + 1$ and $r = \text{select}_1(A_{last}, z + s)$, 
where \chgred{$z = \text{rank}_1(A_{last}, y-1)$} and $s = \text{rank}_c(A_{label}, i)$.

\chgred{Algorithm~\ref{alg:get_children} gives the pseudo-code of \textbf{Children$(i)$}.}

\begin{algorithm}[t]
  \caption{Children algorithm for retrieving the range of children for the entry at position $i$ in the XBW representation.}
  \label{alg:get_children}
  \begin{algorithmic}[1]
  \Function{Children}{$i$}
    \State $c \gets \text{access}(A_{label}, i)$ \Comment{Label of the potential parent}
    \State $y \gets F(c)$ \Comment{First position whose ancestor prefix starts with label $c$}
    \State \chgred{$z \gets \text{rank}_1(A_{last}, y-1)$} \Comment{Number of 1s \chgred{before} position $y$}
    \State $s \gets \text{rank}_c(A_{label}, i)$ \Comment{Number of occurrences of label $c$ up to position $i$}
    \State $l \gets \text{select}_1(A_{last}, z + s - 1) + 1$ \Comment{Start of children block}
    \State $r \gets \text{select}_1(A_{last}, z + s)$ \Comment{End of children block}
    \State \Return $[l, r]$ \Comment{Range of child entries}
  \EndFunction
  \end{algorithmic}
\end{algorithm}

\textbf{Example:} \chgred{We apply} \textbf{Children$(i)$} \chgred{with} $i=5$ \chgred{to jXBW} in Figure~\ref{fig:xbw_construction}(IV):
(i) Compute $c = \text{access}(A_{label}, 5) = I$.
(ii) Compute $y = F(I) = 11$, where position $11$ is the first position \chgred{whose} parent \chgred{has} label $I$.
(iii) Compute \chgred{$z = \text{rank}_1(A_{last}, y-1) = \text{rank}_1(A_{last}, 10) = 7$} and $s = \text{rank}_I(A_{label}, i) = \text{rank}_I(A_{label}, 5) = 1$.
(iv) Compute \chgred{$l = \text{select}_1(A_{last}, z + s - 1) + 1 = \text{select}_1(A_{last}, 7) + 1 = 11$} and \chgred{$r = \text{select}_1(A_{last}, z + s) = \text{select}_1(A_{last}, 8) = 12$}.
(v) The children range is $[11, 12]$.

\subsection{RankedChild and CharRankedChild Operations}
Both \textbf{RankedChild$(i, k)$} and \textbf{CharRankedChild$(i, c, k)$} \chgred{use} the starting position $l$ from the range $[l, r] = \textbf{Children$(i)$}$.
\textbf{RankedChild$(i, k)$} is computed as $l + k - 1$.
\textbf{CharRankedChild}$(i, c, k)$ is computed as follows: 
(i) Compute $y =$ \\ $\text{rank}_c(A_{label}, l-1)$ to obtain the number of occurrences of $c$ up to position $l-1$; 
(ii) Compute $p = \text{select}_c(A_{label}, y + k)$ to find the \chgred{position of the} $(y+k)$-th occurrence of $c$ in $A_{label}$.

\subsection{Parent Operation}
\textbf{Parent$(i)$}: \chgred{This operation leverages} the lexicographic \chgred{order} of the XBW representation. \chgred{Because nodes with the same parent label are stored contiguously, we can find the start of the group of positions that contains $i$ and shares its parent label, count the sibling blocks completed before $i$ in this group, and use this count to locate the parent's position.} \chgred{The operation proceeds as follows:}
(i) Compute $s = \text{rank}_1(A_{diff}, i)$, where $A_{diff}$ is a binary array: $A_{diff}[i]=1$ iff \chgred{the first labels (parent labels) of $A_{anc}[i]$ and $A_{anc}[i-1]$ differ} for $i \neq 1$; $A_{diff}[1] = 1$. For example\chgred{,} in Figure~\ref{fig:xbw_construction}, \chgred{$A_{diff} = [1, 1, 0, 0, 0, 1, 1, 0, 1, 1, 1, 0]$}.
The value $s$ \chgred{is} the number of distinct \chgred{parent labels} up to position $i$.
(ii) Compute $y = \text{select}_1(A_{diff}, s)$, the starting position of the group in $A_{label}$ \chgred{whose ancestor} label sequences in $A_{anc}$ \chgred{share the first label of the sequence} at position $i$.
(iii) Compute $c = \text{access}(A_{pf}, i)$\chgred{,} the label of the parent \chgred{of the} node at position $i$.
(iv) Compute \chgred{$k = \text{rank}_1(A_{last}, i-1) - \text{rank}_1(A_{last}, y-1)$}, the number of sibling \chgred{blocks} with parent label $c$ completed between positions $y$ and \chgred{$i-1$}.
(v) \chgred{Compute} the parent \chgred{position} $p = \text{select}_c(A_{label}, k+1)$, the position of the $(k+1)$-th occurrence of label $c$ in $A_{label}$ (the $+1$ accounts for the parent position relative to the sibling \chgred{block} count).
(vi) Return $p$\chgred{.}

\chgred{Algorithm~\ref{alg:get_parent} gives the pseudo-code of \textbf{Parent$(i)$}.}

\begin{algorithm}[t]
  \caption{Parent algorithm for retrieving the parent position of the entry at position $i$ in the XBW representation.}
  \label{alg:get_parent}
  \begin{algorithmic}[1]
  \Function{Parent}{$i$}
    \If{\chgred{$i = 1$}}
      \State \Return \textsc{NULL} \Comment{Root node has no parent}
    \EndIf

    \State $s \gets \text{rank}_1(A_{diff}, i)$ \Comment{Number of distinct \chgred{parent labels} up to $i$}
    \State $y \gets \text{select}_1(A_{diff}, s)$ \Comment{Start position of the group containing $i$}
    \State $c \gets \text{access}(A_{pf}, i)$ \Comment{Parent label of entry at position $i$}

    \State \chgred{$k \gets \text{rank}_1(A_{last}, i-1) - \text{rank}_1(A_{last}, y-1)$} \Comment{\chgred{Completed sibling blocks}}
    \State $p \gets \text{select}_c(A_{label}, k + 1)$ \Comment{$(k+1)$-th occurrence of label $c$}

    \State \Return $p$ \Comment{Parent position in $A_{label}$}
  \EndFunction
  \end{algorithmic}
\end{algorithm}

\textbf{Example:} \chgred{We apply} \textbf{Parent$(i)$} \chgred{with} $i=5$ \chgred{to jXBW} in Figure~\ref{fig:xbw_construction}(IV):
\chgred{(i) Compute $s = \text{rank}_1(A_{diff}, i) = \text{rank}_1(A_{diff}, 5) = 2$.}
\chgred{(ii) Compute $y = \text{select}_1(A_{diff}, s) = \text{select}_1(A_{diff}, 2) = 2$.}
(iii) Compute $c = \text{access}(A_{pf}, i) = \text{access}(A_{pf}, 5) = A$.
(iv) Compute \chgred{$k = \text{rank}_1(A_{last}, i-1) - \text{rank}_1(A_{last}, y-1)$} \\
\quad\quad\chgred{$= \text{rank}_1(A_{last}, 4) - \text{rank}_1(A_{last}, 1) = 2 - 1 = 1$.}
(v) Compute $p = \text{select}_A(A_{label}, k+1) = \text{select}_A(A_{label}, 2) = 9$.
(vi) Return $p = 9$\chgred{,} the position \chgred{of} the parent of the node at position $i = 5$.

\subsection{SubPathSearch Operation}

\textbf{SubPathSearch$(P)$}: \chgred{This operation narrows} the range of positions in $A_{label}$ \chgred{one} label \chgred{at a time along} the path $P = \langle p_1,p_2,\ldots,p_k \rangle$. Starting with nodes that have the first label $p_1$, we refine the search range to find nodes reachable through \chgred{$P$}. \chgred{The operation proceeds as follows:}
(i) If the path $P$ is empty, return the range \chgred{$[1, |MT|]$}.
(ii) For the first label $p_1$, compute the initial range $[\text{first}, \text{last}]$ using \chgred{$F$}: $\text{first} = F(p_1)$ and $\text{last} = F(p_1 + 1) - 1$, where $F(c)$ gives the first position of nodes whose parent has label $c$.
(iii) For each subsequent label $p_i$ ($i = 2, 3, \ldots, |P|$), refine the current range $[\text{first}, \text{last}]$ by:
(a) Computing $k_1 = \text{rank}_{p_i}(A_{label}, \text{first} - 1)$ and $k_2 = \text{rank}_{p_i}(A_{label}, \text{last})$ to count occurrences of $p_i$ within the current range.
(b) Computing the positions $z_1 = \text{select}_{p_i}(A_{label}, k_1 + 1)$ and $z_2 = \text{select}_{p_i}(A_{label}, k_2)$ to find the first and last occurrences of $p_i$ within the range.
(c) If $i = |P|$ (final label), set $\text{first} = z_1$ and $\text{last} = z_2$ and terminate.
(d) Otherwise, update the range to the children of the matched nodes: $\text{first} = \textbf{RankedChild}(z_1, 1)$ and $\text{last} = \textbf{RankedChild}(z_2, \text{degree}(z_2))$. Here, $\text{degree}(z_2)$ is the number of children of the node at position $z_2$, computed as $r - l + 1$ where $[l,r]$ is the range \chgred{returned} by \textbf{Children$(z_2)$}.
(iv) Return the final range $[\text{first}, \text{last}]$ \chgred{of the} positions of nodes reachable through $P$.

\chgred{Algorithm~\ref{alg:subpath_search} gives the pseudo-code of \textbf{SubPathSearch$(P)$}.}

\begin{algorithm}[t]
  \caption{SubPathSearch algorithm for finding nodes that match a path $P = \langle p_1,p_2,\ldots,p_k \rangle$ on jXBW.}
  \label{alg:subpath_search}
  \begin{algorithmic}[1]
  \Function{SubPathSearch}{$P$}
    \If{$P$ is empty}
      \State \Return \chgred{$[1, |MT|]$} \Comment{Match all entries}
    \EndIf

    \State $first \gets F(p_1)$
    \State $last \gets F(p_1 + 1) - 1$
    \If{$first > last$}
      \State \Return \textsc{NULL} \Comment{No match found}
    \EndIf

    \For{$i = 2$ to $|P|$}
      \State $c \gets p_i$
      \State $k_1 \gets \text{rank}_c(A_{label}, first - 1)$
      \State $k_2 \gets \text{rank}_c(A_{label}, last)$
      \If{$k_2 \leq k_1$}
        \State \Return \textsc{NULL} \Comment{No match in range}
      \EndIf

      \State $z_1 \gets \text{select}_c(A_{label}, k_1 + 1)$
      \State $z_2 \gets \text{select}_c(A_{label}, k_2)$

      \If{$i = |P|$}
        \State \Return $[z_1, z_2]$ \Comment{Final range}
      \EndIf

      \State $first \gets \text{RankedChild}(z_1, 1)$
      \State $last \gets \text{RankedChild}(z_2, \text{degree}(z_2))$

      \If{$first = \textsc{NULL}$ or $last = \textsc{NULL}$}
        \State \Return \textsc{NULL} \Comment{Invalid child range}
      \EndIf
    \EndFor

    \State \Return $[first, last]$
  \EndFunction
  \end{algorithmic}
\end{algorithm}

\section{Helper Algorithms for Substructure Search}
\label{sec:helper_algorithms}

This appendix \chgred{details} the helper \chgred{functions} used by the SubstructureSearch algorithm \chgred{on jXBW}.

\subsection{CompAncestors Algorithm}

\begin{algorithm}[t]
  \caption{CompAncestors algorithm for collecting ancestor positions $|P|-1$ levels \chgred{above the} positions in \chgred{the} range $[first, last]$ for path $P$ on jXBW.}
  \label{alg:compute_ancestors}
  \begin{algorithmic}[1]
  \Function{CompAncestors}{$[first, last], P$}
    \State $ancestor\_set \gets$ empty set
    \State $path\_length \gets |P|$
    
    \For{each position $i$ from $first$ to $last$}
      \State $pos \gets i$
      \For{$j = 1$ to $path\_length - 1$} \Comment{Trace back $|P|-1$ levels}
        \State $pos \gets \Call{Parent}{pos}$
      \EndFor
      \If{$pos$ is not in $ancestor\_set$}
        \State $ancestor\_set \gets ancestor\_set \cup \{pos\}$ \Comment{Add new ancestor}
      \EndIf
    \EndFor
    
    \State \Return $ancestor\_set$
  \EndFunction
  \end{algorithmic}
\end{algorithm}

\subsection{CollectPathMatchingIDs Algorithm}

\begin{algorithm}[t]
  \caption{CollectPathMatchingIDs algorithm for collecting tree identifiers from \chgred{the leaves reached} from \chgred{the candidate} root position $root\_pos$ \chgred{along the} root-to-leaf paths \chgred{of query tree $Q$}.}
  \label{alg:collect_path_matching_ids}
  \begin{algorithmic}[1]
  \Function{CollectPathMatchingIDs}{$root\_pos$, $paths$}
    \State $sets\_of\_ids \gets$ empty list
    
    \For{each $path$ in $paths$}
      \State $path\_leaf\_ids \gets$ empty set
      \State $current\_positions \gets \{root\_pos\}$ \Comment{Start with matching subtree root position}
      
      \For{each $label$ in $path$}
        \State $next\_positions \gets$ empty set
        \For{each $current$ in $current\_positions$}
          \State $k \gets 1$
          \While{$\Call{CharRankedChild}{current, label, k} \neq \textsc{NULL}$}
            \State $child \gets \Call{CharRankedChild}{current, label, k}$
            \State $next\_positions \gets next\_positions \cup \{child\}$
            \State $k \gets k + 1$
          \EndWhile
        \EndFor
        \State $current\_positions \gets next\_positions$
        \If{$current\_positions$ is empty}
          \State \textbf{break} \Comment{No matching children found}
        \EndIf
      \EndFor
      
      \Comment{Collect tree IDs from all matching leaf positions}
      \For{each $leaf\_pos$ in $current\_positions$}
        \State $leaf\_ids \gets A_{ids}[\text{rank}_1(A_{leaf}, leaf\_pos)]$
        \State $path\_leaf\_ids \gets path\_leaf\_ids \cup leaf\_ids$
      \EndFor
      
      \If{$path\_leaf\_ids$ is empty}
        \State \Return $\emptyset$ \Comment{Early termination if any path fails}
      \EndIf
      
      \State $sets\_of\_ids \gets sets\_of\_ids \cup \{path\_leaf\_ids\}$
    \EndFor

    \State \Return $sets\_of\_ids$
  \EndFunction
  \end{algorithmic}
\end{algorithm}

\section{Case Study Details: Chemical Compound Analysis with GPT}
\label{sec:case_study_details}

This appendix \chgred{fully describes} the case study summarized in Section~\ref{sec:case_study}.

To evaluate our jXBW-based substructure search in a real-world scientific context, we conducted a case study on the PubChem dataset (Table~\ref{tab:jsonl_analysis}). \chgred{The query tree}
$$\{\text{"structure"}: \{\text{"atoms"}: [\{\text{"symbol"}: \text{"N"}, \text{"charge"}: 1\}]\}\}$$
\chgred{targets} compounds containing cationic nitrogen atoms (N$^+$).
The query retrieved 838 compounds, \chgred{from} which \chgred{we} randomly sampled \chgred{10} for downstream analysis.

\chgred{We} then asked \chgred{GPT-4} to analyze \chgred{the} physicochemical traits \chgred{of these compounds} and propose possible pharmacological properties. It identified recurring \chgred{structural motifs} such as quaternary or pyridinium nitrogen centers, porphyrin-like aromatic macrocycles, polar functional groups (carboxyl, hydroxyl), and coordination with transition metals (Mn$^{3+}$, Ni$^{3+}$). Based on these motifs, GPT-4 hypothesized potential applications including antimicrobial activity, photosensitization in photodynamic therapy, and DNA/RNA binding.

Figure~\ref{fig:chemical_examples} shows two representative compounds: a Mn$^{3+}$-coordinated porphyrin derivative with carboxyphenyl groups (left) and a pyridinium-substituted porphyrin complex (right). Both structures exemplify the combination of N$^+$-based charge centers, extended conjugation, and potential redox activity that GPT-4 identified as key structural motifs.

This case study demonstrates the synergy between fast substructure search and large language models, showing how \chgred{integrating} symbolic querying with neural reasoning can \chgred{speed up compound triage, the generation of structure--function hypotheses, and drug discovery}.

\fi

\end{document}